\documentclass[namedreferences]{solarphysics}
\usepackage[optionalrh]{spr-sola-addons} 
\usepackage{graphicx}        
\usepackage{color}           
\usepackage{url}             
\usepackage[outdir=./]{epstopdf}

\newcommand{\Rs}{$ R_{\odot}$}

\newcommand{\DEM}{cm$^{-5}$ K$^{-1}$}
\newcommand{\EM}{cm$^{-5}$}
\newcommand{\DN}{DN cm$^{5}$ $s^{-1}$ pix$^{-1}$}

\begin{document}

\begin{article}

\begin{opening}

\title{SITES: Solar Iterative Temperature Emission Solver for differential emission measure inversion of EUV observations}

\author{Huw~\surname{Morgan}}
\author{James~\surname{Pickering}}
\runningauthor{Morgan \& Pickering}
\runningtitle{SITES: Solar Iterative Temperature Emission Solver}

   \institute{$^{1}$ Physics Department, Aberystwyth University, Ceredigion, Cymru, SY23 3BZ
                     email: \url{hmorgan@aber.ac.uk} \\ 
             }

\begin{abstract}
Extreme UltraViolet (EUV) images of the optically-thin solar corona in multiple spectral channels give information on the emission as a function of temperature through differential emission measure (DEM) inversions. The aim of this paper is to describe, test, and apply a new DEM method named the Solar Iterative Temperature Emission Solver (SITES). The method creates an initial DEM estimate through a direct redistribution of observed intensities across temperatures according to the temperature response function of the measurement, and iteratively improves on this estimate through calculation of intensity residuals. It is simple in concept and implementation, is non-subjective in the sense that no prior constraints are placed on the solutions other than positivity and smoothness, and can process a thousand DEMs per second on a standard desktop computer. The resulting DEMs replicate model DEMs well in tests on Atmospheric Imaging Assembly (AIA) synthetic data. The same tests show that SITES performs less well on very narrow DEM peaks, and should not be used for temperature diagnostics below $\sim$0.5MK in the case of AIA observations. The SITES accuracy of inversion compares well with two other established methods. A simple yet powerful new method to visualise DEM maps is introduced, based on a fractional emission measure (FEM). Applied to a set of AIA full-disk images, the SITES method and FEM visualisation show very effectively the dominance of certain temperature regimes in different large-scale coronal structures. The method can easily be adapted for any multi-channel observations of optically-thin plasma and, given its simplicity and efficiency, will facilitate the processing of large existing and future datasets.
\end{abstract}
\end{opening}

\section{Introduction}
\label{intro}

Understanding the physics of the Sun's atmosphere demands increasingly detailed and accurate observations. The development of new analysis methods to gain physical observables from remote sensing observations is an ongoing and critically important effort. As part of this effort, this paper presents a new Differential Emission Measure (DEM) method for the temperature/density analysis of solar coronal optically-thin emission lines. The Extreme UltraViolet (EUV) spectrum from the solar atmosphere contains several strong emission lines from highly-ionised species above a relatively low background. These lines are emitted from the hot corona only, thus narrowband EUV observations are an excellent probe of the low corona, with little contamination from the underlying photosphere and lower atmosphere. 

The concept of using EUV line intensities to estimate the temperature of the emitting plasma is based on the temperature of formation of the line: a range of temperatures at which a certain ion can exist, and the relative population of that ion as a function of temperature. Thus calibrated observations of two lines with different formation temperatures can give a constraint on the dominant plasma temperature. Based on this concept, the simplest approach to estimating a dominant coronal temperature is the line ratio method, which assumes an isothermal plasma (see, for example, the description and criticism of \opencite{weber2005}). 

In the general case, imaging instruments provide an observed intensity integrated across a narrow bandpass that spans one or more spectral line - this is the case for an EUV imaging instrument such as AIA. Thus the temperature response of each channel may be computed based on the wavelength response of that channel, and modelled line intensities from an established atomic database (such as Chianti, \cite{dere1997}) using certain assumptions (e.g. Maxwell-Boltzmann distributions and thermal equilibrium). The measured intensity of multiple bandpasses, or channels, with different temperature responses, allow the estimation of emission as a function of temperature, or a DEM. A DEM is a powerful characterisation of the coronal plasma - it is an estimate of the total number of electrons squared along the observed line of sight (similar to a column mass) at a given temperature. The DEM method has revealed the general temperature characteristics of the main structures seen in the corona: for example, closed-field active regions are hot and multithermal ($>$2MK), open-field regions are colder ($<$1.1MK), and in between is the quiet corona ($\sim$1.4MK) \cite{delzanna2013,hahn2011,mackovjak2014,hahn2014}. Changes in DEM over time are related to heating or cooling, and can be applied over large datasets to reveal solar cycle trends \cite{morgan2017}. 

For an imaging instrument such as AIA, the DEM method inverts measured intensities in a small number of bandpasses to give the emission as a function of temperature across a large number of temperature bins. This is an underdetermined problem that requires additional constraints on the solution, such as positivity and smoothness. There are several types of DEM methods in use, well summarized in the introduction to \opencite{hannah2012}. One method is that of \opencite{hannah2012}, which uses Tikhonov regularization to find an optimal weighting between fitting the data and satisfying additional constraints of positivity of the DEM (negative emission is unphysical), minimising the integrated emission, and smoothness of the result. To our knowledge, the most computationally fast method is that of \opencite{cheung2015}, based on Simplex optimization of a set of smooth basis functions, or a sparse matrix. \opencite{plowman2013} use a parametric functional form for the DEM, solved with a regularized inversion combined with an iterative scheme for removal of negative DEM values. A similar parametric form is also used by \opencite{nuevo2015} in the context of coronal tomography and a localised DEM.

This work presents a new DEM inversion method in section \ref{method}. The method is introduced in the context of the type of imaging observations made by an instrument such as AIA, but can easily be generalised to any observation where the measurement temperature response is known. Tests of the method on synthetic observations made from model DEMs are made in section \ref{synthetic}, along with a non-rigorous test on computation time. Section \ref{aiadata} discusses uncertainty in AIA measurements, and applies the method to data. An effective method to visualise DEM maps is also presented in section \ref{aiadata}. A brief summary is given in section \ref{summary}.

\section{The DEM method} 
\label{method}

\begin{figure}    
   \centerline{\includegraphics[width=0.95\textwidth,clip=]{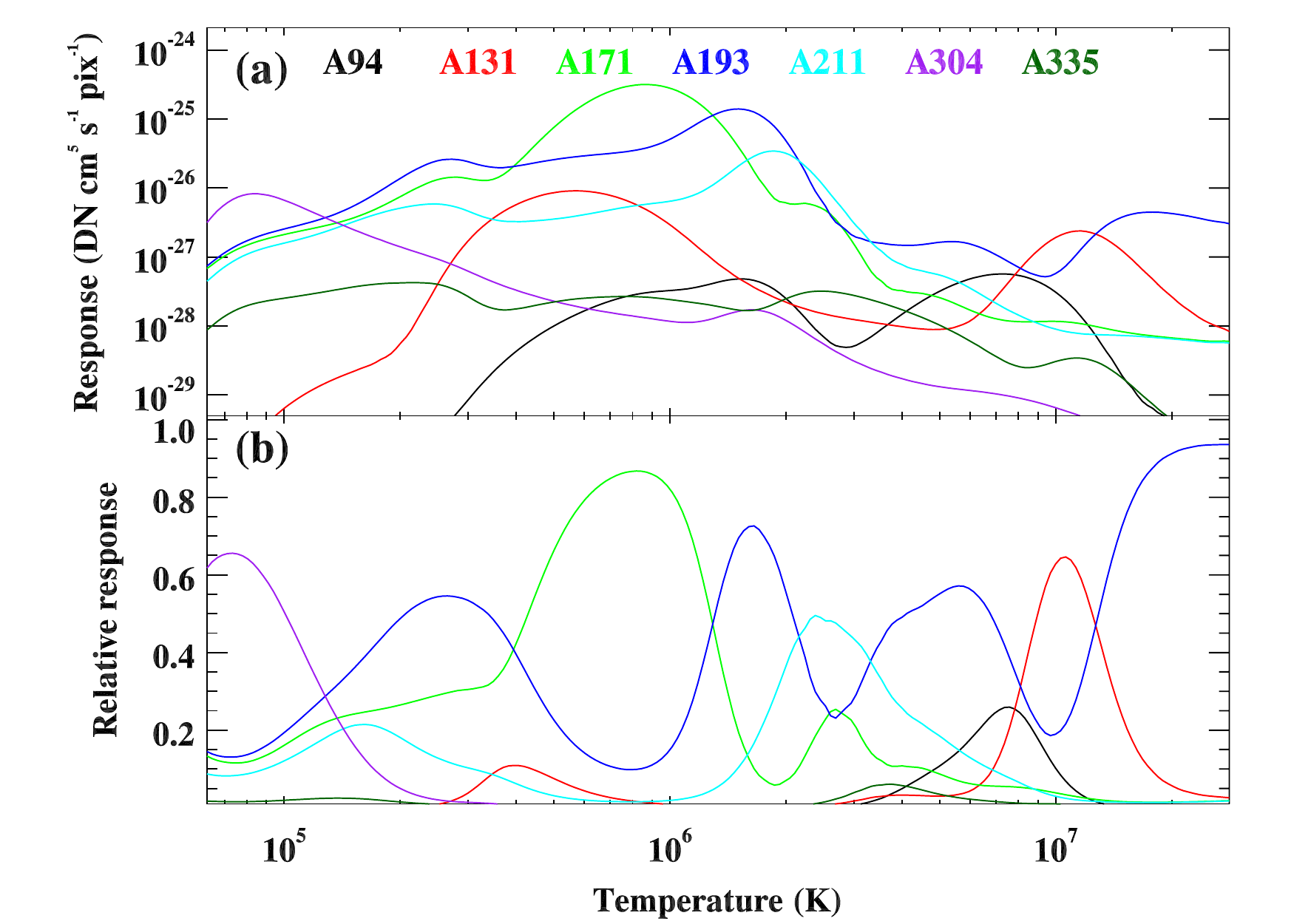}}
   \caption{(a) The temperature response of the seven AIA EUV channels, as given by the standard AIA calibration routines in Solarsoft, based on Chianti atomic data and normalised through cross-calibration with EVE data. This set is for date 2011/01/01. (b) The relative response as a function of temperature. At a given temperature, the relative responses sum to unity over all channels.}
    \label{response}
  \end{figure}

A set of intensities $I_0, I_1, ...I_{n-1}$ are measured by $i=0,1,...,n-1$ AIA channels, with associated errors  $\sigma_i$. Each channel's response as a function of temperature, $R_{ij}$, is known for a set of temperature bins indexed $j=0,1,...,n_t-1$. This work uses the response functions as given by the standard AIA Solarsoft routines, calculated from the Chianti atomic database \cite{dere1997,landi2012}, cross-calibrated over time with EVE observations and including a correction to the 94\AA\ channel calibration \cite{boerner2014}. An example of these functions are shown in figure \ref{response}a. We assume that each response function has a relative error $\varepsilon_i$, constant over all temperature. These uncertainties are discussed in a following section.

Before considering weightings associated with the relative noise in each channel, we first introduce the simple concept of relative temperature responses. The relative response for a given channel (indexed $i$) and temperature (indexed $j$), $S_{ij}$, is calculated as
\begin{equation}
\label{relres}
S_{ij}=\frac{R_{ij}}{\sum_{i=0}^{n-1}R_{ij} }, 
\end{equation}
so that, at a given temperature bin, the relative responses sum to unity over all channels. The relative responses are an useful value, and are shown in figure \ref{response}b. For example, at very high temperatures ($>$10MK), the relative response of the 193 channel is almost 1, showing that any DEM method using solely AIA data will be very uncertain at these flare temperatures, since only one channel is giving information at this temperature. A similar argument holds for low ($<$0.1MK) temperatures, where the relative response of the 304 channel increases to 0.7. The relative response (further weighted by the relative error in each channel, to be introduced later) is used in the DEM method to combine the information from each channel. Thus, as can be seen from figure \ref{response}b, at temperatures near 0.9MK, the resulting DEM will be dominated by the 171 channel. 

It is convenient to include information on the relative measurement error of each channel, and the estimated errors of the response functions in the relative response. Thus equation \ref{relres} becomes
\begin{equation}
\label{relres2}
S_{ij}=\frac{w_i R_{ij}}{\sum_{i=0}^{n-1}w_i R_{ij} }, 
\end{equation}
where $w$ is a weighting based on the measurement and calibration errors, 
\begin{equation}
\label{w}
w_i=\frac{1}{\sqrt{(\frac{\sigma_i}{I_i})^2 + \varepsilon_i^2}}.
\end{equation}
Thus channels with smaller relative errors (i.e. higher signal to noise, and/or less uncertainty in response function) will have greater weighting in the DEM estimate.

An initial DEM estimate, $D_j$ is given by 
\begin{equation}
\label{dem1}
D_j=\left[ \sum_i^{n-1} S_{ij} \left( I_i \frac{R_{ij} \Delta T_j}{\sum_j^{n_t-1} (R_{ij} \Delta T_j)^2 }  \right)   \right] \otimes K,
\end{equation}
where $K$ is a smoothing kernel. In words, a set of $n$ DEM profiles is calculated, one for each channel, based directly on the response function of each channel (the expression within the round brackets). Since the observed intensity is distributed over the DEM temperature range according to the response function of that channel, integrating these individual DEMs over temperature would result in exactly the observed intensities. These individual DEMs are combined into a single DEM through a weighted mean, using the weighting of the relative responses (i.e. product with the relative responses, $S_{ij}$, and summation over $i$). This DEM is convolved with a narrow Gaussian kernel over temperature $K$, to ensure a smooth DEM. The kernel $K$ is a Gaussian profile in logarithmic temperature, with a width (standard deviation) of 3.2 bins in logarithmic temperature, for 43 temperature bins over a temperature range of 0.07-20MK. These values are found through trial and error, with the criteria that the smoothing width is kept at a minimum value whilst still resulting in smooth DEMs. The width of the smoothing kernel is, in fact, the only subjective choice in this procedure. If more temperature bins are set, then the width of the kernel should be increased in proportion.

From this initial DEM, a set of modelled intensities $M_i$ is computed for each channel by
\begin{equation}
\label{iexp}
M_i = \sum_j^{n_t-1}D_j R_{ij} \Delta T_j.
\end{equation}
The residual, or difference between the observed and modelled intensities, is calculated as $I_i^\prime=I_i - M_i$. This residual intensity is fed back into equation \ref{dem1} (taking the place of $I_i$ in the equation), and the resulting residual DEM added to the previous DEM. At this step, the main DEM is thresholded to a minimum value of zero since the residuals may result in a negative DEM at certain temperatures - thus a positivity constraint is applied. This process is iterated until convergence is reached, defined as when the weighted mean of the absolute ratios between the measurement residuals at the current iteration and the initial measurement, drops below an appropriately small threshold, for example 1\%. The weights for this mean are those given by equation \ref{w}. This is a sensible criteria for convergence - the process stops when the mean changes to the output DEM become small, with weighting towards the higher certainty measurements. There are similarities in this iterative approach to that of \opencite{plowman2013}, which computes residual data intensities at several iterations in order to adjust the estimated DEM and eliminate negative intensities. However, the core DEM estimation at each iteration, given by equation \ref{dem1}, is quite different to their method.

An estimate for the DEM error $d$ at each temperature bin $j$ is
\begin{equation}
\label{err}
d_{j} = \sqrt{ \sum_i^{n-1} S_{ij} \left[ \left( \sigma_i/I_i \right) ^2 + \varepsilon_i^2 \right] }.
\end{equation}
The relative measurement error, $\sigma_i/I_i$, and the response function relative uncertainty $\varepsilon_i$, are summed in quadrature, giving the total squared measurement error for each channel. These are multiplied by the relative response $S_{ij}$ in order to distribute over temperatures, and summed over all channels, corresponding to the equivalent steps in the DEM estimate of equation \ref{dem1}. The square root of this value gives the final DEM uncertainty. A complete error propagation treatment should consider the smoothing kernel and multiple iterations, but these steps would defeat the aim of implementing an efficient method. The uncertainties given by the simple calculation of equation \ref{err} give values that agree well with tests involving varying the input measurements according to measurement noise, as is shown in subsection \ref{robustnoise}.

\section{Demonstration using synthetic data}
\label{synthetic}

\subsection{A simple test}
\label{simpletest}
A model DEM is produced by the Gaussian
\begin{equation}
\label{gauss}
D^\prime = A \exp \left( \left[ \frac{t-t_c}{w_t} \right]^2 \right),
\end{equation}
with peak maximum $A=1.4\times 10^{21}$\DEM, central temperature $t_c =$1.4MK, and width $w_t=0.9$MK. Using the AIA response functions (as shown in figure \ref{response}), synthetic observations are created for the 7 channels, in units of \DN\ by integrating the product of the DEM with the response functions over temperature. Measurement uncertainties are given by the AIA Solarsoft routine aia\_bp\_estimate\_error. The synthetic observations are input into SITES, using 43 temperature bins within a temperature range of 0.07-20MK, with a regular bin size in logarithmic temperature. The method terminates at 101 iterations when convergence, as defined in the method, reaches 1\%. To avoid edge effects caused by the smoothing truncation, the first and last DEM bins are discarded, leaving 41 temperature bins in the results. The maximum absolute relative difference between input target intensities and the method's derived intensities (the $M_i$ of equation \ref{iexp}) is 3.5\%\ for the 335 channel. The mean absolute measurement difference $T_I$ over all channels is defined as
\begin{equation}
\label{measdev}
T_I = \frac{1}{n} \sum_i^{n-1}  \frac{| I_i - M_i | }{I_i},
\end{equation}
and is 1.2\%\ for this simple test. 

\begin{figure}    
   \centerline{\includegraphics[width=0.95\textwidth,clip=]{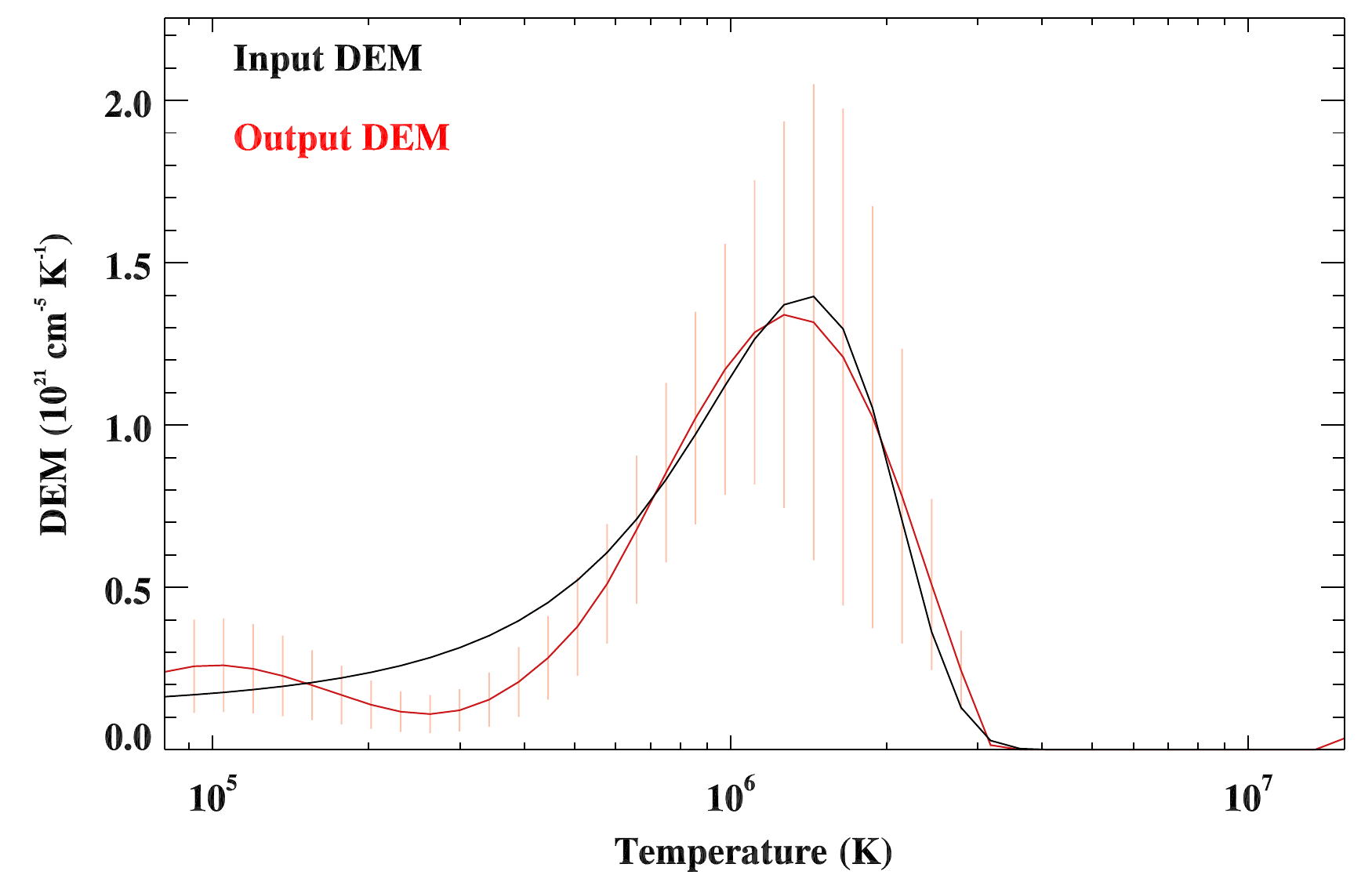}}
   \caption{Comparing input (black) and output (red) DEM curves for the simple case of a single Gaussian in temperature (equation \ref{gauss}). The light red error bars show the uncertainty in the fitted DEM.}
    \label{simpledem}
  \end{figure}

The resulting DEM is compared to the target input DEM in figure \ref{simpledem}. The median absolute relative DEM deviation $T_D$, between the input DEM $D^\prime$ and output estimated DEM $D$, is defined as 
\begin{equation}
\label{demdev}
T_D =  \mathrm{median} \frac{| D_j^\prime - D_j | }{D_j^\prime},
\end{equation}
and is 26\%\ for this simple test. The median is calculated over temperature bins $j$ and is used here rather than the mean to avoid the influence of very small, or zero, values of input DEM at some temperatures bins. 

The correlation $c$ between the input and output DEM curves is defined as
\begin{equation}
\label{corr}
c= \frac{ \sum_j^{n_t-1}  ( D_j - \bar{D_j} ) ( D_j^\prime - \bar{D_j^\prime} ) } {\sqrt{ \sum_j^{n_t-1}  ( D_j - \bar{D_j} )^2 \sum_j^{n_t-1}  ( D_j^\prime - \bar{D_j^\prime} )^2   }}
\end{equation}
and is 98\% for this simple test. Thus the position and width of the main peak is well fitted. 

One concern is the range of temperatures used for the calculation. The examples here have temperatures limited to between 0.07 and 20MK. Some channels have significant values in their response functions outside this range, leading to an inherent uncertainty that can be included in the estimate of each response function's uncertainty. Thus an estimate of the relative uncertainty in the response of each channel is given by
\begin{equation}
\varepsilon_i = \sqrt {e_i^2 +\left( \frac{\sum_{j0} R_{ij}\Delta T_j}{\sum_{j1} R_{ij}\Delta T_j}\right)^2},
\end{equation}
where $e_i$ is the calibration uncertainty for each channel, and the subscript $j1$ are the indices of temperature bins included within the temperature range, and $j0$ otherwise. The $e_i$ are given by AIA Solarsoft routines, and is 50\% for the 94, 131 and 304 channels, 25\% otherwise. $\varepsilon_i$ is a large uncertainty, ranging from 27\%\ for the 171 channel, to 103\%\ for the 131 channel. Channels with large contributions to their response functions outside of the temperature range of interest have a lesser weighting in calculating the final DEM. 

\begin{figure}    
   \centerline{\includegraphics[width=0.95\textwidth,clip=]{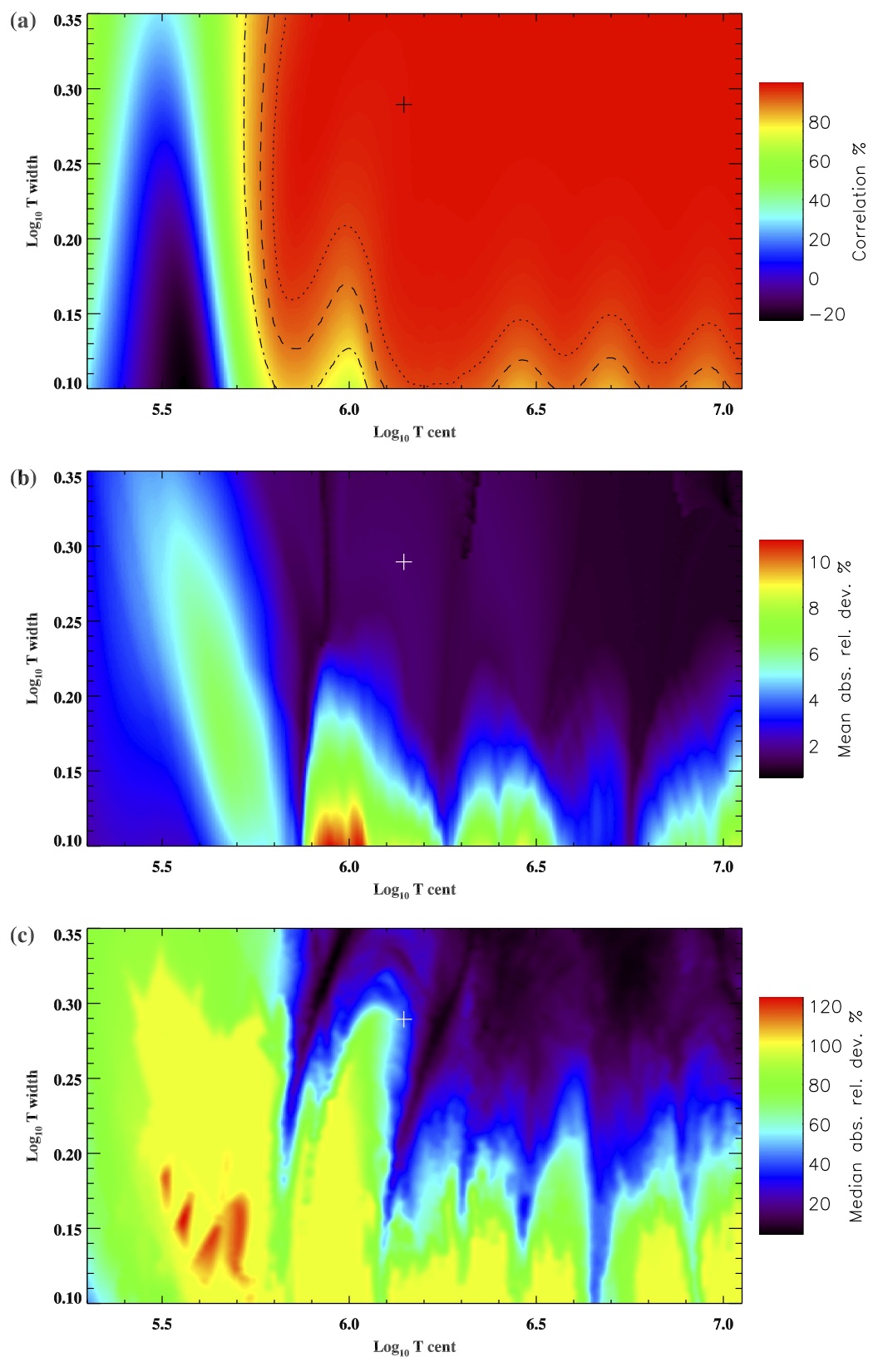}}
   \caption{(a) Correlation $c$ between input and SITES-inverted DEM profiles, (b) mean absolute relative deviation $T_I$ between input measurement and output fitted measurement and (c) median absolute relative deviation $T_D$ of input and SITES-inverted DEM profiles. These are calculated for a range of centers and widths in logarithmic temperature of single-Gaussian DEM profiles. The dotted, dashed and dot-dashed lines in (a) show the 95, 90 and 80\% correlation levels respectively. The cross symbol shows the position corresponding to the single-Gaussian example shown in figure \ref{simpledem}.}
    \label{simpleparams}
  \end{figure}

The single-Gaussian DEM is used as a test of SITES across a broad range of Gaussian central temperature and Gaussian widths. The central temperature is increased from $\log T$ 5.3 to 7.05 in 160 increments, and the widths from $\log T$ 0.1 to 0.35 in 160 increments (note this differs from the example of figure \ref{simpledem}, which is formed from Gaussians in linear temperature). For each input DEM, synthetic measurements are calculated and given as input to SITES, as above for figure \ref{simpledem}. The correlation between input and output DEM, $c$, as given by equation \ref{corr} is calculated, giving a measure of the similarities of the profiles. This is shown in figure \ref{simpleparams}a. A broad range of central temperatures and widths bounded by the dotted line give correlations above 95\%. Poor correlations, below 80\%, are found for low temperatures below $\log T$ 5.7, and for very narrow profiles at all central temperatures. Figure \ref{simpleparams}b shows the mean absolute relative deviation of the input and output measurements, $T_I$, as given by equation \ref{measdev}. The worst match, at close to 10\% deviation, is found for low temperatures or higher temperatures at narrow widths. The deviation otherwise is good, with the majority of the parameter space at values of 4\% or lower. This is to be expected, given that the iterative scheme is designed to reduce this deviation. Figure \ref{simpleparams}c shows the median absolute relative deviation between the input and output DEMs, $T_D$, as given by equation \ref{demdev}. For the broad region dominated by very high correlations, the deviation is around 15-50\%. This deteriorates to over 50\% for low temperatures below $\log T$ 5.7, or for narrow profiles at all temperatures. 

In summary, SITES performs poorly for narrow DEM profiles at all temperatures. This is inherent to estimating DEMs from an instrument such as AIA, regardless of the method, given the broad multiple-peaked temperature profiles in most channels. SITES performs very poorly for DEMs peaked at cool temperatures below $\log T$ 5.7 ($\sim$0.5MK). At higher temperatures, and broader peaks, SITES performs very well, with 95\% correlation with the target input DEMs.

\subsection{A complex test}
\label{complextest}

\begin{figure}    
   \centerline{\includegraphics[width=0.95\textwidth,clip=]{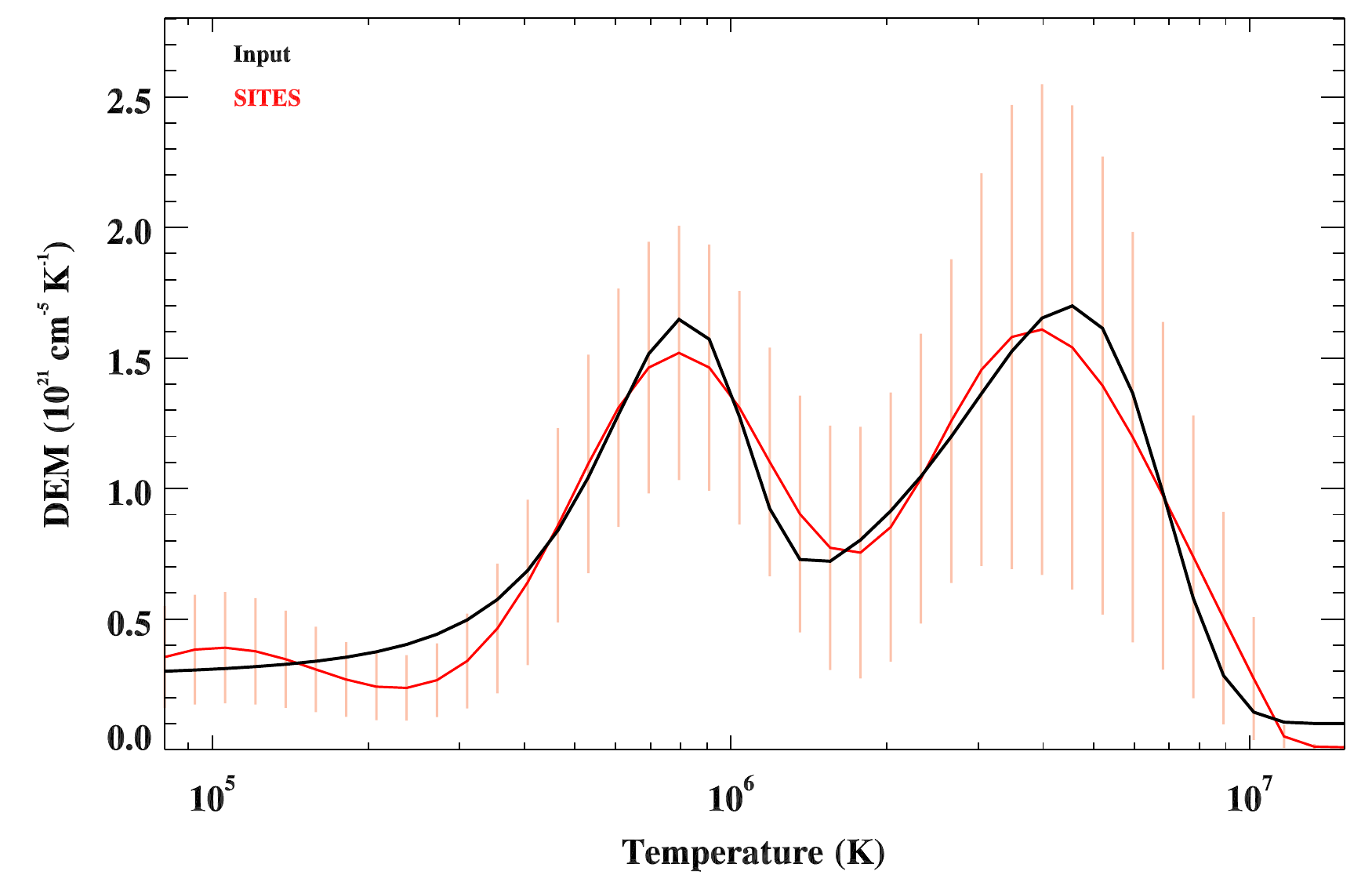}}
   \caption{Comparing input (black) and output (red) DEM curves for the complex case of two Gaussians in temperature and a constant background. The light red error bars show the uncertainty in the output DEM.}
    \label{complexdem}
  \end{figure}

A more complex model DEM sums 2 Gaussian peaks over a constant background. The background emission has a value of $10^{20}$\DEM, and the Gaussian peaks have amplitude $A= [1.6,1.6] \times 10^{21}$\DEM, centered at temperatures $t_c=[0.8,4.5]$ MK, with widths $w_t=[0.35,3.0]$ MK. Synthetic observations are created from this model DEM as for the simple case above. The comparison between input and output DEM is shown in figure \ref{complexdem}. The number of iterations is 114, and $T_I$ is 1.3\%, with a maximum deviation of 3.4\%\ for the 94 channel. $T_D$ is 12\%. The position of the 2 peaks in the resulting DEM estimate agree well with the input DEM - the method is effective at finding these peaks, reflected in the $c=97.7$\%\ correlation between input and output DEM.

\begin{figure}    
   \centerline{\includegraphics[width=0.9\textwidth,clip=]{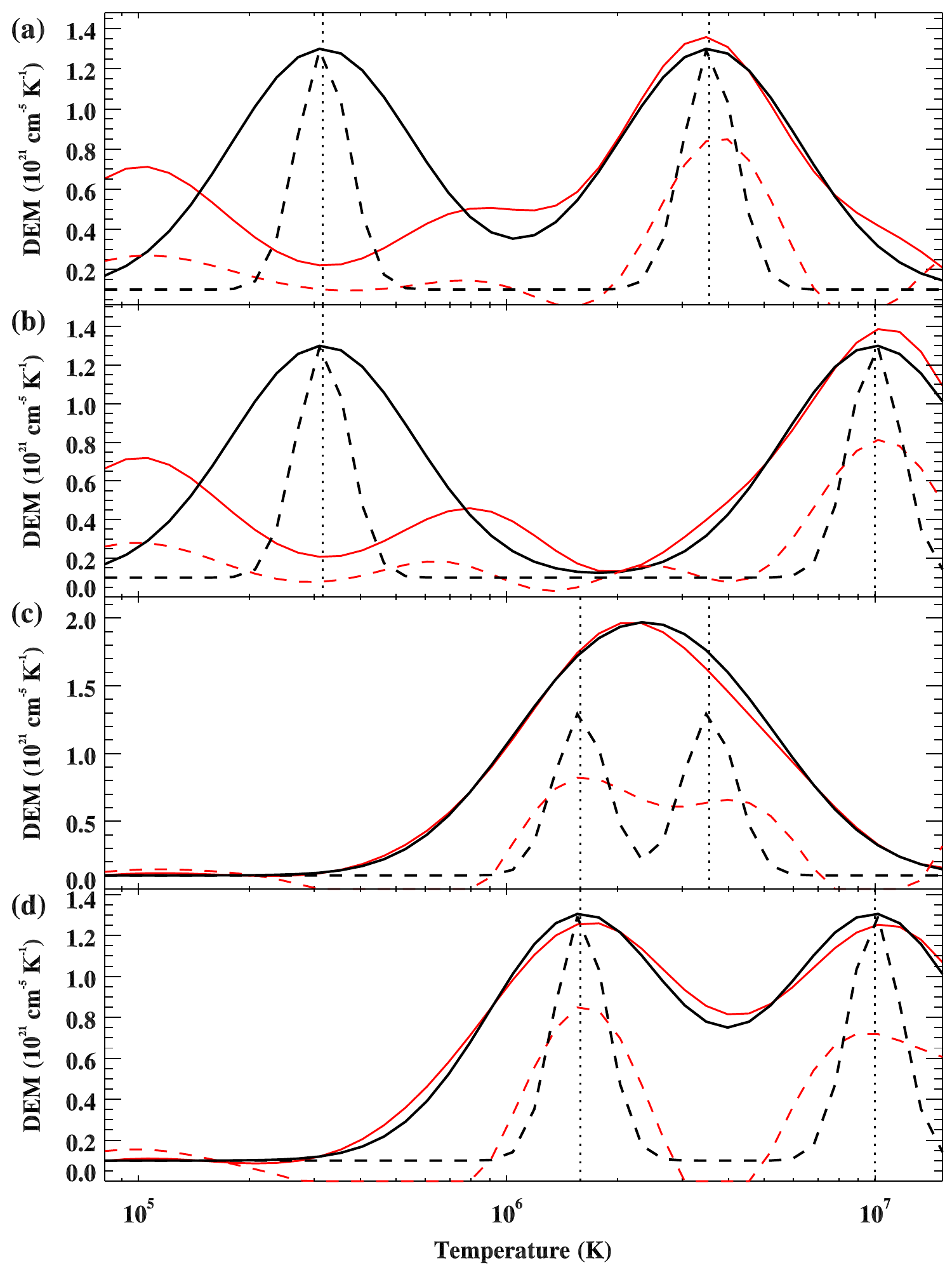}}
   \caption{DEM profiles formed from two Gaussians in logarithmic temperature plus a constant background. The black lines are the input DEM, the red lines are the SITES DEM. The solid (dashed) lines are for wide (narrow) Gaussian profiles (0.35 and 0.1 in logarithmic temperature respectively). Four examples are shown here for the logarithmic peak temperatures of (a) 5.5 and 6.55, (b) 5.5 and 7.0, (c) 6.2 and 6.55, and (d) 6.2 and 7.0. The vertical dashed lines show the central temperature of each peak. }
    \label{cp_examples}
  \end{figure}

The performance of SITES is tested for various combinations of parameters for the two-Gaussian plus background DEM profile through varying the central temperature of each peak. This experiment is repeated for two cases of wide and narrow Gaussians. The Gaussians are formed in logarithmic temperature (note this differs from the example of figure \ref{complexdem}, which is formed from Gaussians in linear temperature). Figure \ref{cp_examples} shows 4 characteristic examples of the two-Gaussian parameter space. Figure \ref{cp_examples}a is for a cool peak at $\log T = 5.5$ and a hot peak at $\log T = 6.55$. For the wide Gaussian profiles (solid lines), the hot peak is well-fitted by SITES, but the fit for the cool peak is poor. The same holds for the narrow Gaussians (dashed lines). The position of the hot narrow peak is found by SITES, although the method struggles to fit the profile properly, with regions next to the peak at zero emission. Figure \ref{cp_examples}b is for a cool peak at $\log T = 5.5$ and a hot peak at $\log T = 7.0$, with similar results to \ref{cp_examples}a. Figure \ref{cp_examples}c is for a cool peak at $\log T = 6.2$ and a hot peak at $\log T = 6.55$, thus the wide Gaussians are blended. SITES fits this profile very well. There are two closely-placed yet distinct peaks in the narrow Gaussian DEM. The SITES DEM also shows two peaks, but is far smoother than the target DEM. Note also the tendency for regions close to the two peaks to have zero emission. Figure \ref{cp_examples}d is for a cool peak at $\log T = 6.2$ and a hot peak at $\log T = 7.0$. Similar to the previous case, the wide DEMs are fitted very well. The narrow peaks are found by SITES, but are smoother, and tend to zero in nearby regions.

\begin{figure}    
   \centerline{\includegraphics[width=0.95\textwidth,clip=]{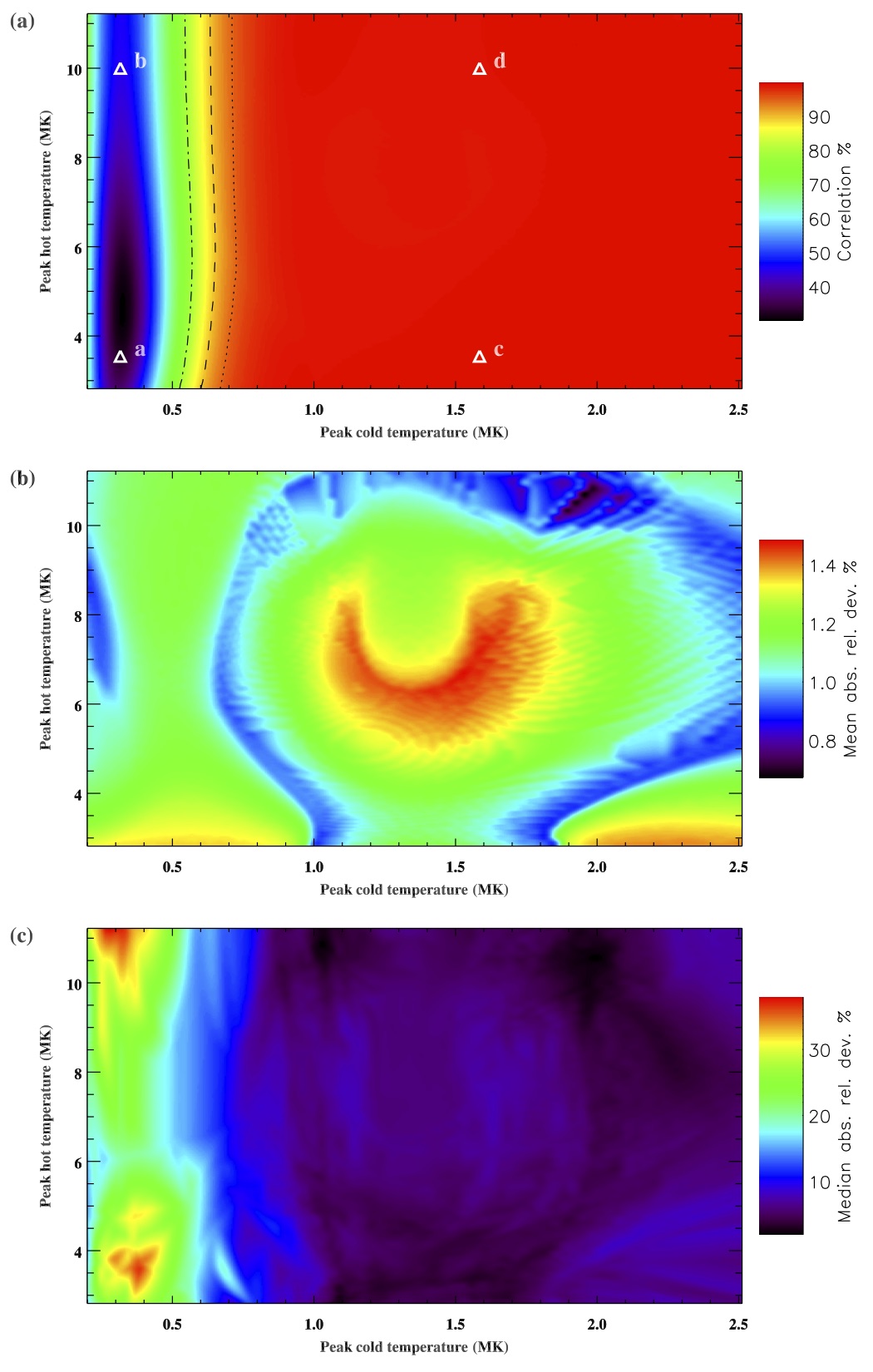}}
   \caption{(a) Correlation $c$ between input and SITES-inverted DEM profiles, (b) mean absolute relative deviation $T_I$ between input measurement and output fitted measurement and (c) median absolute relative deviation $T_D$ of input and SITES-inverted DEM profiles. These are calculated for a range of central peak temperatures for two wide Gaussian DEM profiles, with the $x$-axis ($y$-axis) corresponding to the central temperature of the cooler (hotter) peak. The four triangle symbols labelled a-d in (a) correspond to the four example profiles of figure \ref{cp_examples}a-d. The dotted, dashed and dot-dashed lines in (a) show the 95, 90 and 80\% correlation levels respectively.}
    \label{cp_wide}
  \end{figure}

  Figure \ref{cp_wide} shows the performance of SITES for a range of central temperatures for both Gaussian peaks, for the case of the wide Gaussians. The correlation between the input and SITES DEMs, shown in figure \ref{cp_wide}a, is excellent ($c>80$\%) for all peak hot temperatures, and cool temperatures above 0.5MK. Below this cool temperature, the performance of SITES is poor, despite the close fit to the input measurement as shown in figure \ref{cp_wide}b. The same poor fit for low temperatures is seen in the median absolute relative deviation of the input and output DEMs in figure \ref{cp_wide}c. 
  
\begin{figure}    
   \centerline{\includegraphics[width=0.95\textwidth,clip=]{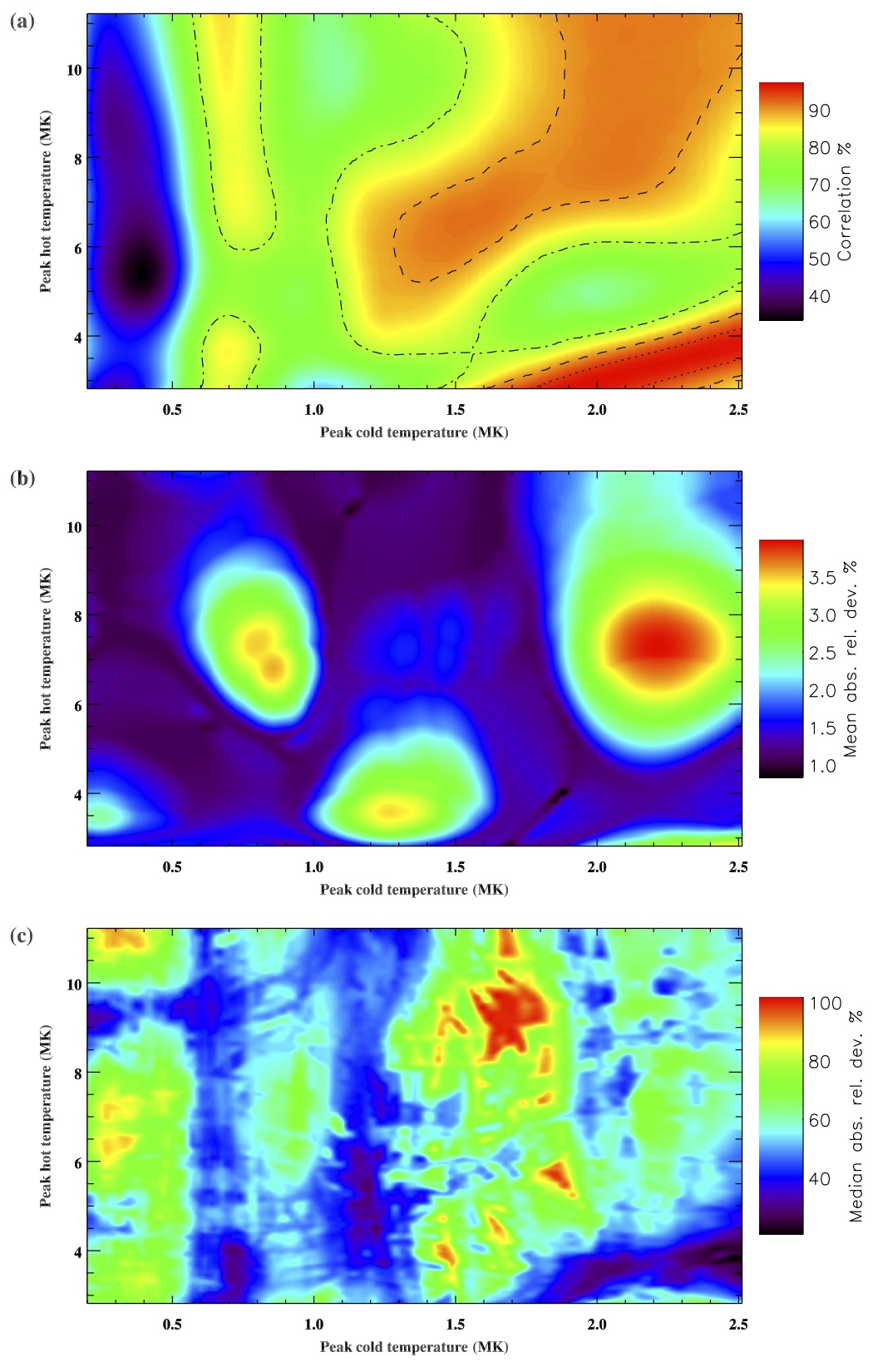}}
   \caption{As figure \ref{cp_wide}, but for the two narrow Gaussians.}
    \label{cp_narrow}
  \end{figure}

  Figure \ref{cp_narrow} shows the same parameter test for narrow Gaussian profiles. Overall, the correlation, fit to measurement, and fit to DEM have deteriorated throughout the parameter space. The very poor fit at low cold peak temperatures remains. In summary, the conclusions for a complex double-Gaussian DEM profile are similar to the case of a single Gaussian in the previous section. SITES performs poorly for narrow DEM profiles at all temperatures, and performs very poorly for DEMs which contain peaks at cool temperatures below $\log T$ 5.7 ($\sim$0.5MK). At higher temperatures, and broader peaks, SITES performs very well, with $c=95$\% correlation with the target input DEMs. 
  
\subsection{Computational speed and convergence threshold}

\begin{figure}    
   \centerline{\includegraphics[width=0.95\textwidth,clip=]{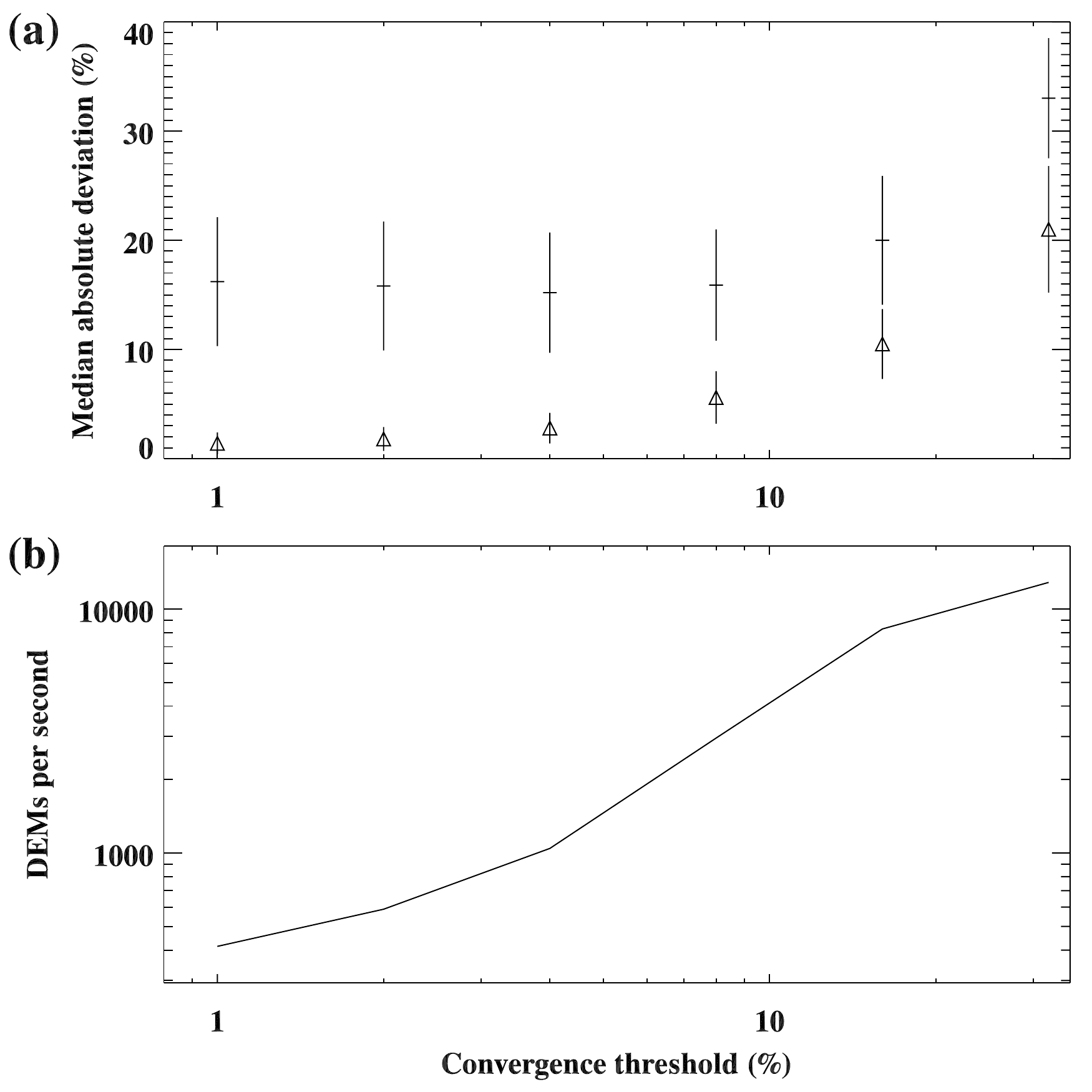}}
   \caption{(a) The percentage median absolute relative deviation of the estimated DEM from the model DEM (crosses) and the relative measurement residuals (triangles) as a function of convergence threshold. These are calculated for a thousand DEMs, with the input measurements varied randomly according to the measurement uncertainty estimates, giving the error bars. (b) The number of DEMs calculated per second as a function of convergence threshold on a standard desktop PC (see text).}
    \label{convergence}
  \end{figure}

To test computational efficiency, SITES is applied 1000 times to the complex 3-Gaussian plus background DEM distribution, imposing a variation on input channel intensities based on their randomisation according to the measurement uncertainty estimates at each run. This experiment is repeated for convergence thresholds of 1, 2, 4, 8, 16 and 32\%. Figure \ref{convergence}a summarises the performance of SITES as a function of the increasing convergence thresholds through the median absolute residuals of the measurements (goodness of fit), and the median absolute deviation of the resulting DEMs compared to the target model DEM. There is no significant deterioration of achieving the target DEM up to the 8\% convergence threshold. The measurement residuals similarly remain small up to the 8\% convergence threshold. On a Linux desktop Intel Core i7-4790 CPU with 16Gb memory the 1000 runs are timed, with the number of DEMs calculated per second shown in figure \ref{convergence}b. Based on these results, for real data we set a convergence threshold of 4\%, which can process around 1000 DEMs per second. This speed is similar to regularized matrix inversion-based methods such as \opencite{hannah2012} or \opencite{plowman2013}.

\subsection{Robustness to noise}
\label{robustnoise}

\begin{figure}    
   \centerline{\includegraphics[width=0.95\textwidth,clip=]{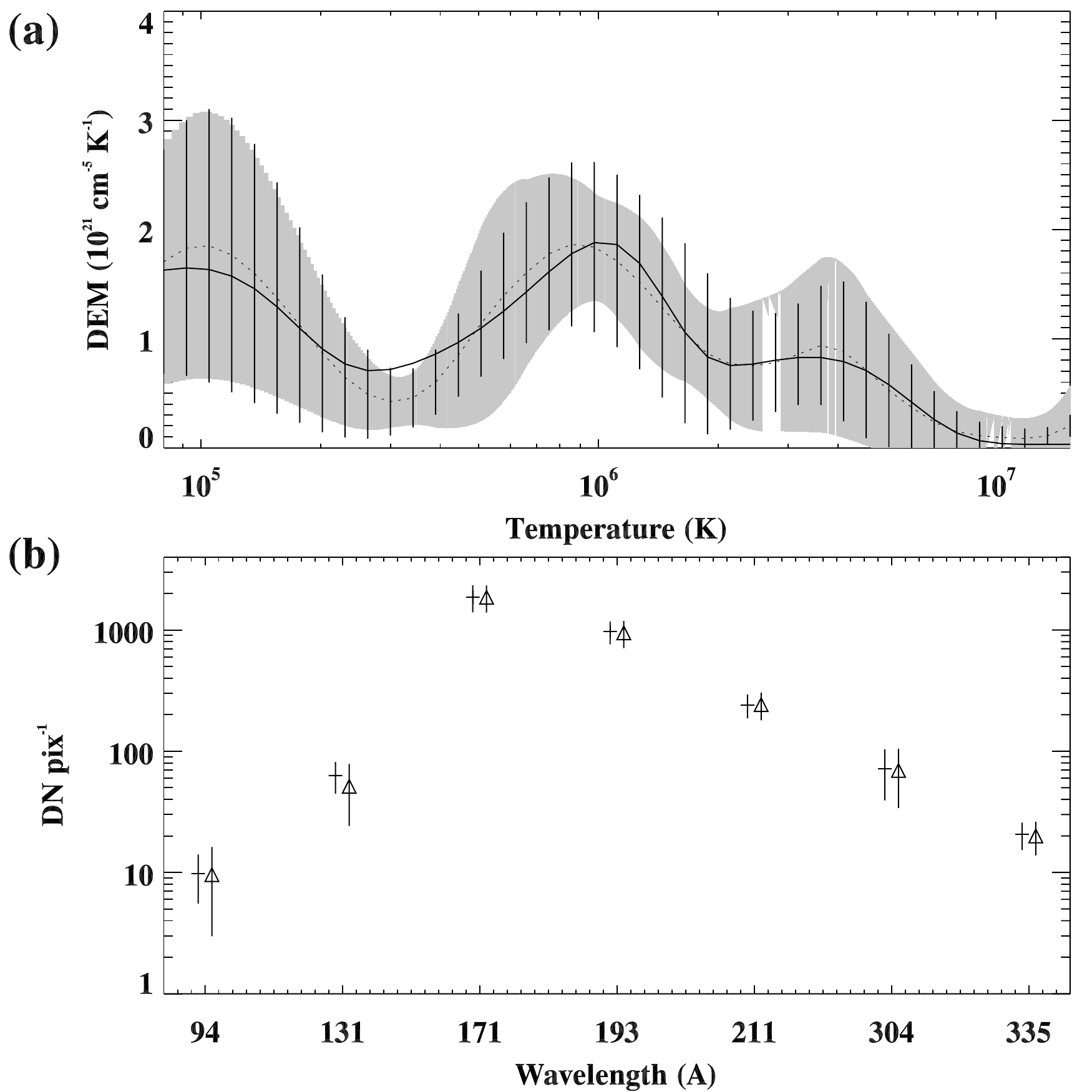}}
   \caption{(a) Applying SITES a thousand times to noise-varying measurements gives a mean DEM (dotted line) and the standard deviation DEM (shaded area) at each temperature bin. The vertical error bars show the estimated error bars gained from equation \ref{err}, averaged over the thousand experiments. The solid black line is the input model DEM (as described in section \ref{complextest}). (b) The triangle symbols show the input measurements in the absence of noise, with the associated error bars showing the noise amplitude in each channel. The cross symbols and associated error bars show the mean and standard deviation fit to the data over the thousand cases (gained from the DEMs using equation \ref{iexp}).}
    \label{robust}
  \end{figure}

This section tests the reliability of SITES in the presence of noise. A complex DEM formed from 3 Gaussian peaks and a constant background is used to create synthetic measurements. This DEM is shown as the solid black line in figure \ref{robust}a. For a thousand repetitions, the measurements are varied randomly according to a noise amplitude given by the measurement and calibration errors, and the resulting DEMs recorded. The convergence factor is set at 4\%, at a value that will typically be used for practical use on real data. 

Figure \ref{robust}a shows the mean DEM, calculated over the thousand repetitions, as a dotted line. This can be compared to the input model DEM which is shown as a bold solid line. The grey shaded region shows the standard deviation of DEMs over the thousand repetitions. The error bars show the mean DEM errors as calculated by equation \ref{err}. Figure \ref{robust}b shows the input measurements in each channel, in the absence of noise, as triangle points, with the error bars showing the noise level. The cross symbols and error bars show the mean and standard deviation of the fitted measurements (i.e. gained from the output DEM through equation \ref{iexp}). Despite the large variations in the DEM values, the 3-peak profile is well replicated. The presence of noise does not lead to DEMs that deviate significantly beyond that expected given the uncertainties. The uncertainty estimate of equation \ref{err} reflects well the true variation of the output DEMs. Integrating the product of the DEMs with the response functions (equation \ref{iexp}) shows that the method is fitting the input data correctly. As can be seen in figure \ref{robust}b, the only systematic discrepancy is seen for the low-signal 131 channel, where the method gives a small positive residual.

\begin{figure}    
   \centerline{\includegraphics[width=0.95\textwidth,clip=]{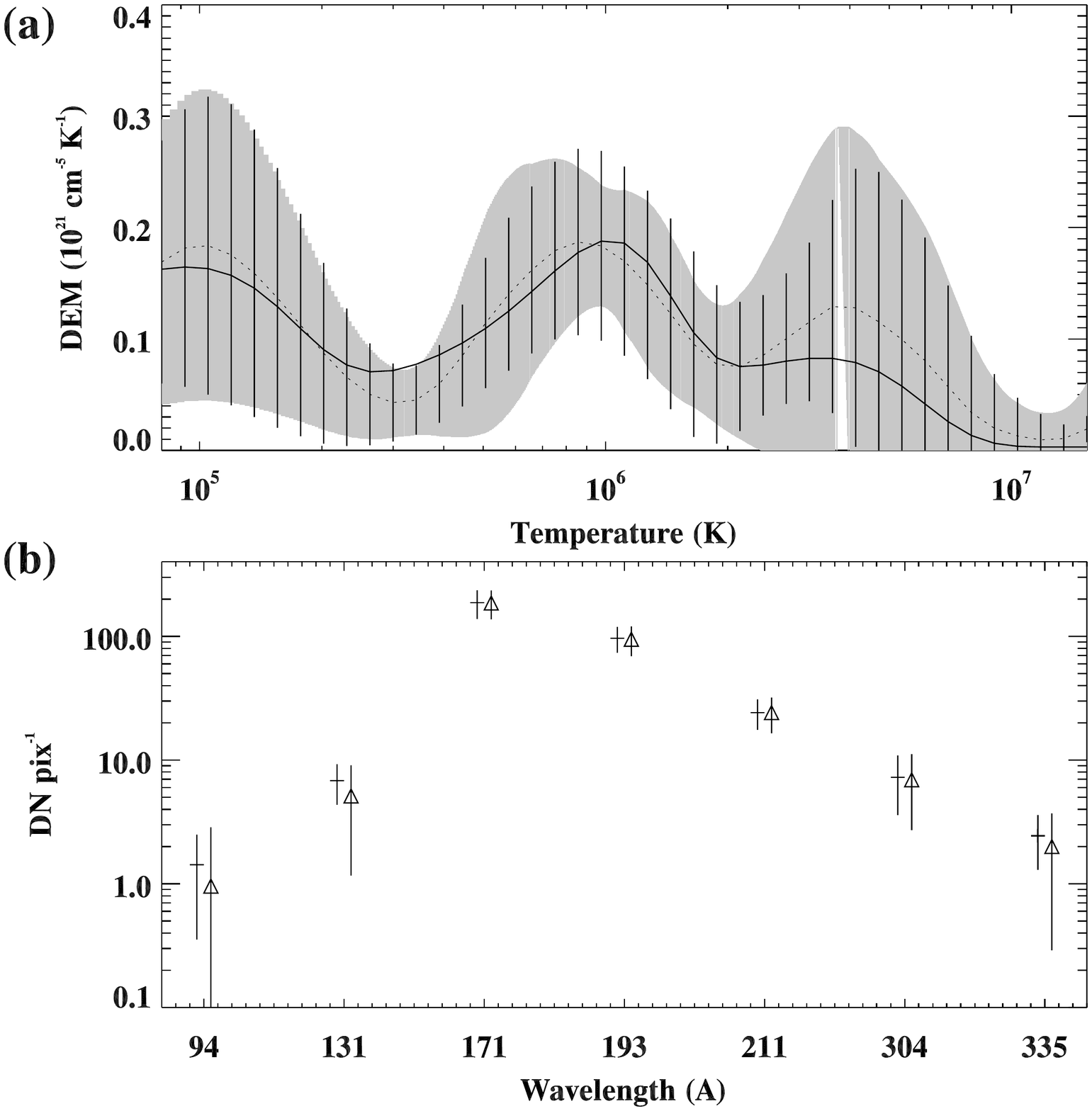}}
   \caption{As figure \ref{robust}, but for the very noisy case of a signal 10 times less intense.}
    \label{robust2}
  \end{figure}

 Figure \ref{robust2}a shows the distribution of DEMs resulting from running the experiment for a signal 10 times lower than the previous example. In this very noisy case, SITES performs reasonably well, although the third DEM peak at high temperature is overestimated. The estimated error bars have increased correctly given the increase in noise across temperatures up to $\sim$2MK. Above this temperature, the uncertainty is underestimated. From figure \ref{robust2}b, the measurement residuals are systematically too high for the lower-signal 94, 131 and 335 channels.

\subsection{Comparison with other methods}

SITES is compared here with the method of \opencite{cheung2015}, hereafter called Sparse Matrix Inversion (SMI), and with the method of \opencite{hannah2012}, hereafter called Tikhonov Regularization (TR). Both the simple single Gaussian DEM of section \ref{simpletest} and the multiple Gaussian plus constant background DEM of section \ref{complextest} are used to create synthetic measurements that are given as input to SITES, SMI and TR. All three methods use identical temperature response functions, measurements and measurement errors for inversion. The TR method is called with the default order equal to zero, and we show the positive-constrained solution.

\begin{figure}    
   \centerline{\includegraphics[width=0.98\textwidth,clip=]{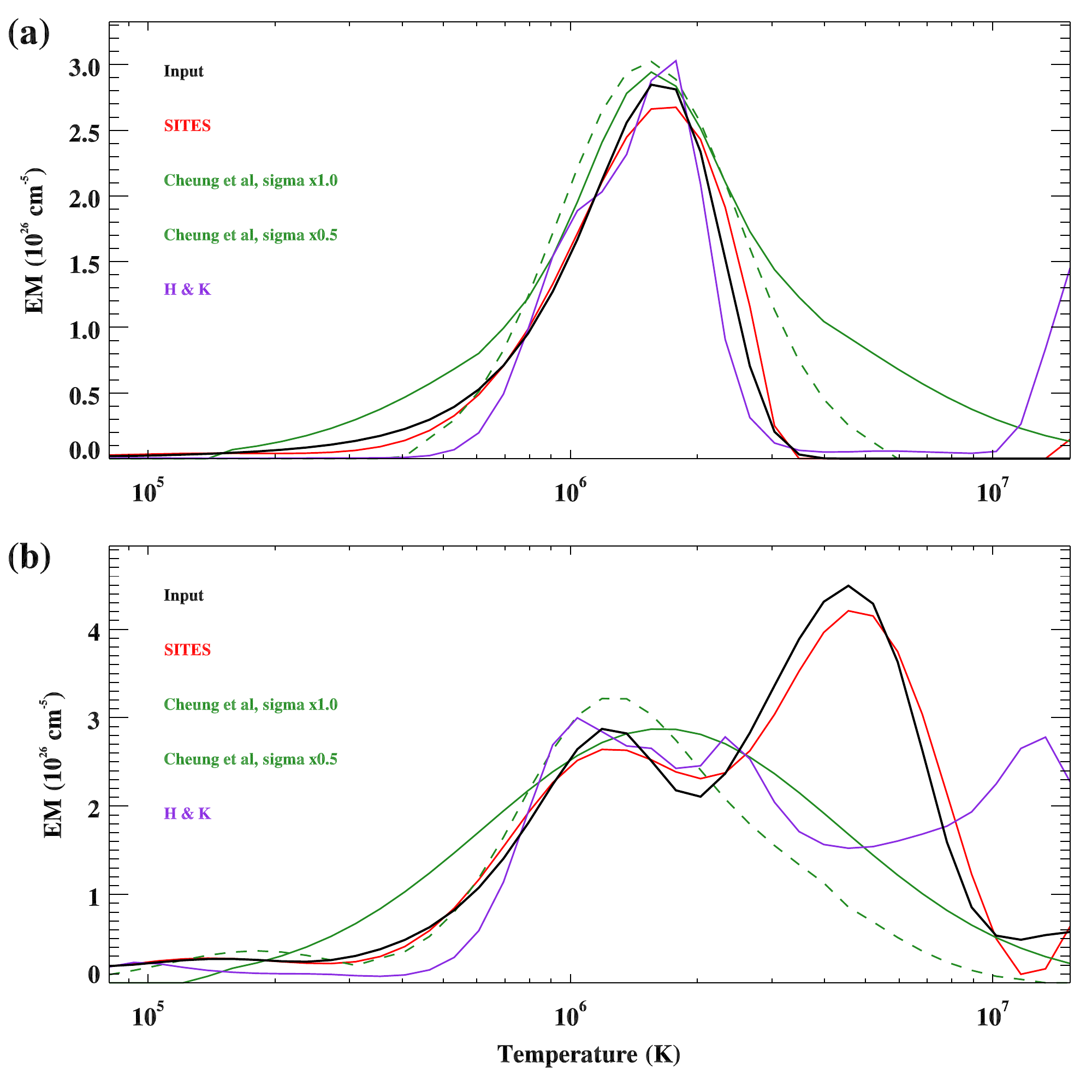}}
   \caption{Comparison of input target emission (black line), SITES (red line with error bars), Cheung \emph{et al} (2015) (SMI, green lines) and Hannah \& Kontar (2012) (blue line with error bars) for (a) the simple single Gaussian DEM of section \ref{simpletest} and (b) the multiple Gaussian DEM of section \ref{complextest}. The SMI method is run for two different values of the width of Gaussian basis functions (see text). Note that these plots show values of emission (EM) rather then DEM, corresponding to the output of the SMI software.}
    \label{cheung}
  \end{figure}

The resulting emissions as functions of temperature for the single Gaussian case is shown in figure \ref{cheung}a. The result for the default choice of the SMI Gaussian basis functions is shown as a solid green line. It is obvious that this choice of basis functions gives an EM result which is too wide. Halving the width of the basis functions (dashed green line) gives a decent fit to the input EM curve, although emission is too high towards the high-temperature wing of the distribution. TR gives a good fit except at the highest range of temperature, where a steep increase is seen. SITES also has a small increase at the highest temperature bin. SITES outperforms both SMI and TR for this example, in closely fitting the Gaussian peak and giving zero DEM at higher temperatures. 

Figure \ref{cheung}b shows the result for a double-Gaussian input DEM. In the case of using the broad (default) SMI basis functions (solid green line), the estimated EM broadly covers the correct temperature region, but fails to identify the individual peaks. The narrow basis functions (dashed green line) successfully identifies the EM peak near $T=1$MK, but fails to invert the other peak, and gives an overall profile which is too narrow across temperature. TR is effective in finding the cooler $T=1$MK peak but fails to identify the main peak near 4MK. SITES outperforms both SMI and TR for the two-Gaussian DEM profiles, in successfully finding all three Gaussian peaks plus the constant background. 

\begin{figure}    
   \centerline{\includegraphics[width=1.01\textwidth,clip=]{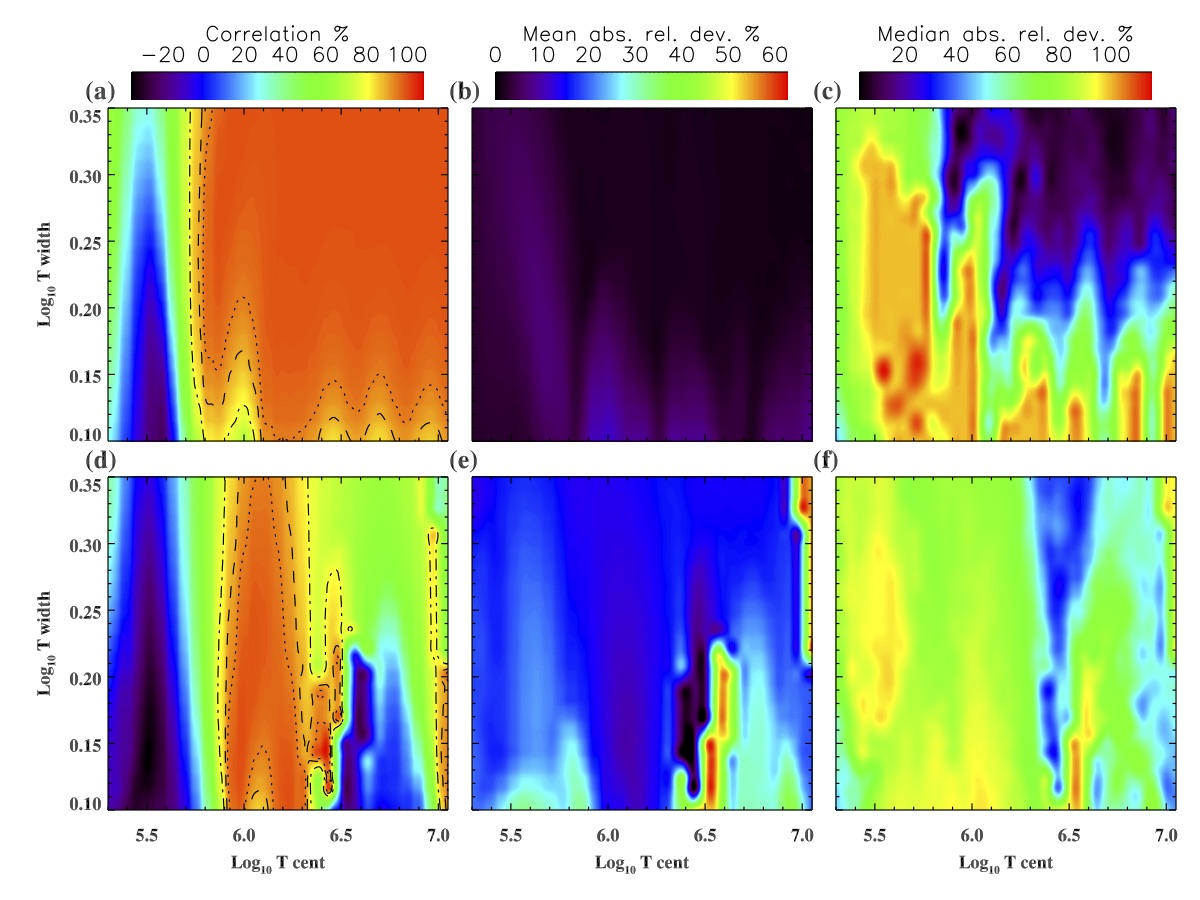}}
   \caption{(a) Correlation $c$ between input and SITES-inverted DEM profiles, (b) mean absolute relative deviation $T_I$ between input measurement and SITES output fitted measurement and (c) median absolute relative deviation $T_D$ of input and SITES-inverted DEM profiles. (d)-(f) Same as (a)-(c), but for the TR method. These are calculated for a range of centers and widths in logarithmic temperature of single-Gaussian DEM profiles. The dotted, dashed and dot-dashed lines in (a) and (d) show the 95, 90 and 80\% correlation levels respectively. The color bars at the top of each column are common to the plots of both methods.}
    \label{comp}
  \end{figure}

The comparison of SITES to the TR method is extended to a parameter search for the case of a single-Gaussian plus background input DEM profile. The parameter space is the same as in section \ref{simpletest}, but with a reduced number of bins (30 bins in Gaussian central temperature and 20 bins in Gaussian width). Results are shown in figure \ref{comp}, with the top row showing the SITES performance (almost identical to figure \ref{simpleparams}, with a different color scale range), and the bottom row showing the TR method performance. The DEM input-output correlation of figures \ref{comp}a and b show a poor inversion for both methods at low temperatures ($\log T < 5.7$). Above this temperature, SITES outperforms TR for almost all central temperatures and widths. SITES also more closely fits the input data by a considerable margin, as shown in figures \ref{comp}b and e. Figures \ref{comp}c and e show the median absolute relative deviation of input and output DEMs for both methods. SITES has a larger region of small deviation ($<40$\%), and more profiles that have a very small deviation ($<20$\%), but also has some regions of higher deviation than TR. In summary, SITES generally gives better perofrmance than TR in this noiseless comparison.

\begin{figure}    
   \centerline{\includegraphics[width=1.01\textwidth,clip=]{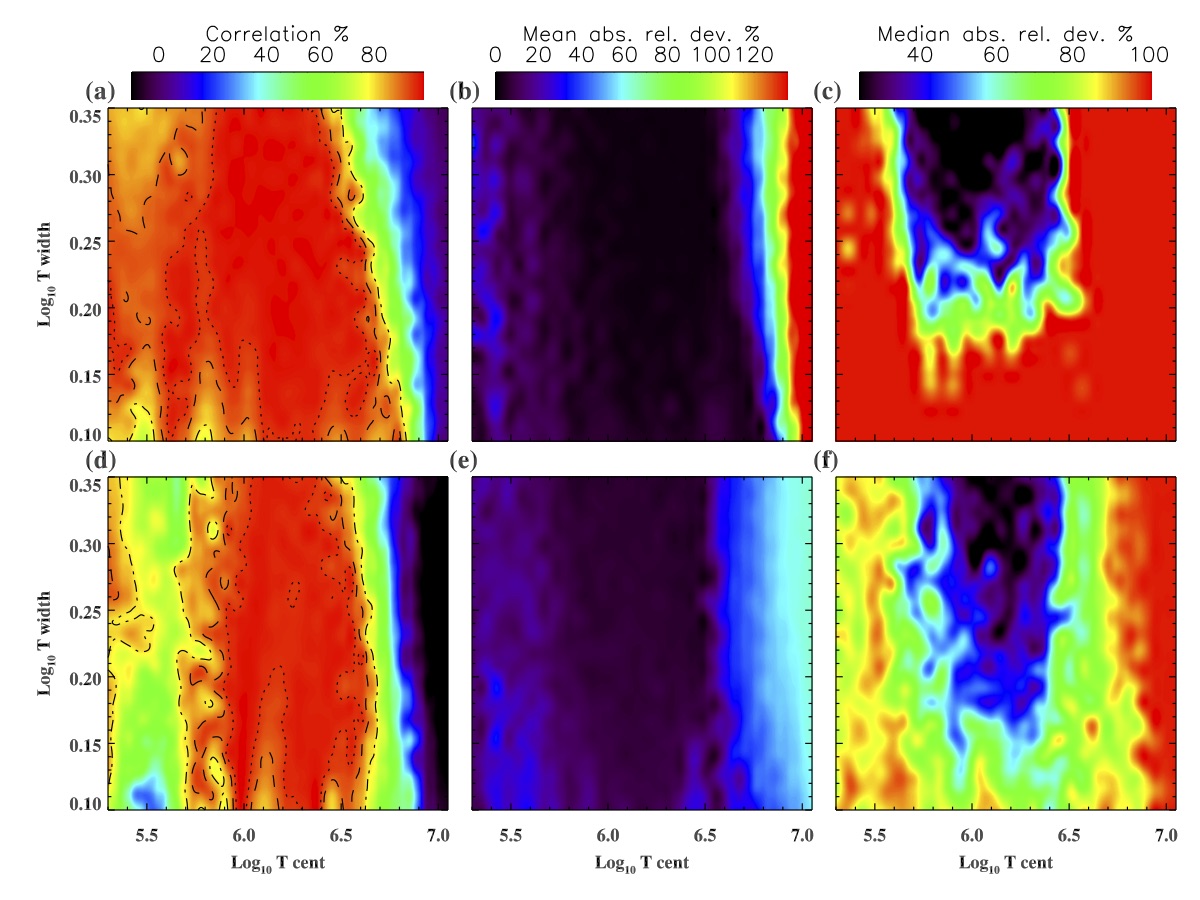}}
   \caption{As figure \ref{comp}, but for the case of input data modulated by Poisson noise. These values show the mean calculated over 15 repetitions, with the intensity values varying randomly with an amplitude set by the Poisson uncertainty.}
    \label{comp2}
  \end{figure}

 As suggested by figure \ref{comp}e, the TR method may be underfitting the data, therefore the comparison with SITES may be unfair since the input data has no randomness associated with noise. This is addressed by repeating the test 15 times, allowing the input data to vary randomly according to a Poisson distribution, comparing the output DEM at each repetition to the input DEM, and taking the mean correlation and measurement/DEM deviations over the 15 cases. To give an idea of the noise amplitude, at a central DEM temperature of $\log T$=6.4 and $\log T$ width 0.26, the relative Poisson noise is 19\% for the lowest signal 94 channel, and 1.4\% for the 193 channel. Results are shown in figure \ref{comp2}. The input/output DEM correlation is generally better for SITES compared to TR (figures \ref{comp2}a and d), whilst the DEM deviation (figures \ref{comp2}c and f) is worse for SITES. In summary, both methods perform similarly for noisy data, with SITES giving an overall better match to the general DEM profiles (a broader region of higher correlation), and TR giving closer absolute values of DEM (a broader region of lower deviation).
  
We note that we have not investigated with any rigour the various parameters of SMI. We have, for example, only used two choices of the basis function widths. We further note that SMI is extremely fast compared to SITES, around a factor of 100 faster depending on the choice of SITES convergence factor. For the TR method, we have experimented with changing the choice of order (which sets the regularization constraints), with similar results to those shown for order equal to zero. At a convergence threshold of 4\%, SITES is of comparable speed to TR. 

\section{Application to AIA data}
\label{aiadata}

\subsection{Data processing and error estimates}
\label{aiaprocessing}

\begin{figure}    
   \centerline{\includegraphics[width=0.95\textwidth,clip=]{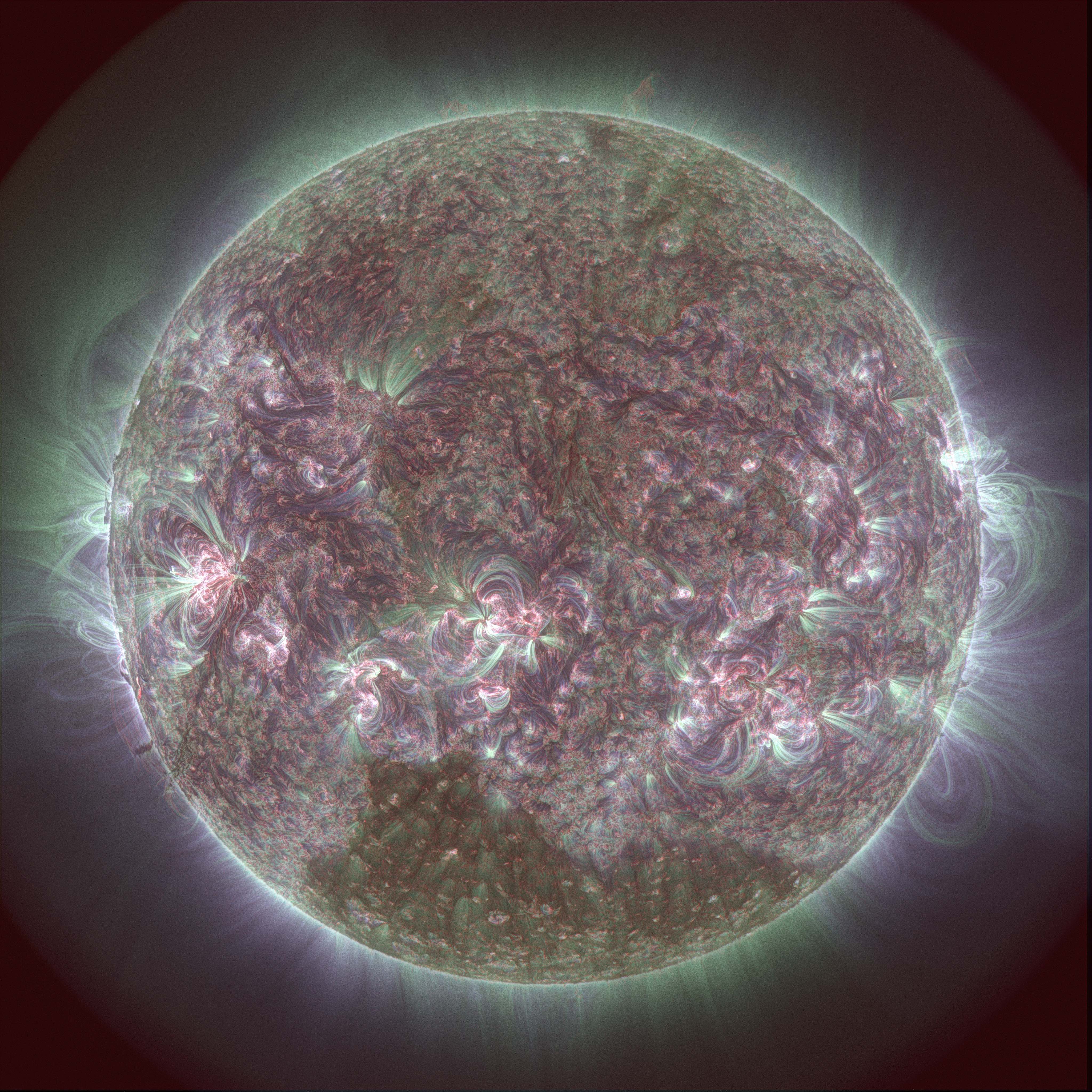}}
   \caption{A context image from 2015/01/01 03:00. All 7 AIA channels contribute to this composite, with the temperature response of each channel between 0.05 and 7.0MK specifying that channel's contribution to the red, green and blue colour channels of the output images. The image is processed with Multiscale Gaussian Normalization to enhance fine-scale structure (Morgan \& Druckmuller, 2014).}
    \label{aiaimage}
  \end{figure}

The standard SDO procedure read\_sdo.pro is used to open a set of full-resolution images in the 7 EUV channels of AIA. An example from 2015/01/01 03:00 is used here. Figure \ref{aiaimage} shows a colour composite processed using Multiscale Gaussian Normalization to provide context \cite{morgan2014}. Each channel's image is shifted in the $x$ and $y$ dimensions so that the central pixel corresponds to the solar disk center, as given by the header image geometry information. A secondary sub-pixel fine alignment is achieved through aligning each image to the 193 channel image, using a phase correlation method to estimate the required shift \cite{druckmuller2009,fisher2008}, and cubic interpolation to apply the shift. For the example set of images, these pixel shifts are listed in table \ref{tabledata}. The mean signal calculated over all pixels on the solar disk is listed for each channel in the table.

\begin{center}
\begin{table}
\begin{tabular}{ c | c | c | c | c  }
{Channel} & {$T (s)$} & {$x_s$} & {$y_s$} & {$\bar{I}$ (DN pix$^{-1}$)}\\
\hline
{94} & {2.9} & {0.86} & {-1.47} & {3}\\
{131} & {2.9} &  {1.59} & {-1.10} & {13}\\
{171} & {2.0} &  {-0.37} & {-0.62} & {271}\\
{193} & {2.0} &  {-       } & {-       } & {411}\\
{211} & {2.9} &  {-0.10} & {0.36} & {207}\\
{304} & {2.9} &  {0.91} & {-0.91} & {28}\\
{335} & {2.9} &  {0.81} & {-0.64} & {5}\\
\end{tabular} 
\caption{Some characteristics of an AIA observation set, with columns showing channel, exposure time, $x$-shift (fine alignment relative to the 193 channel), $y$-shift, and mean intensity (on the disk).}
\label{tabledata}
\end{table}
\end{center}

\begin{figure}    
   \centerline{\includegraphics[width=0.95\textwidth,clip=]{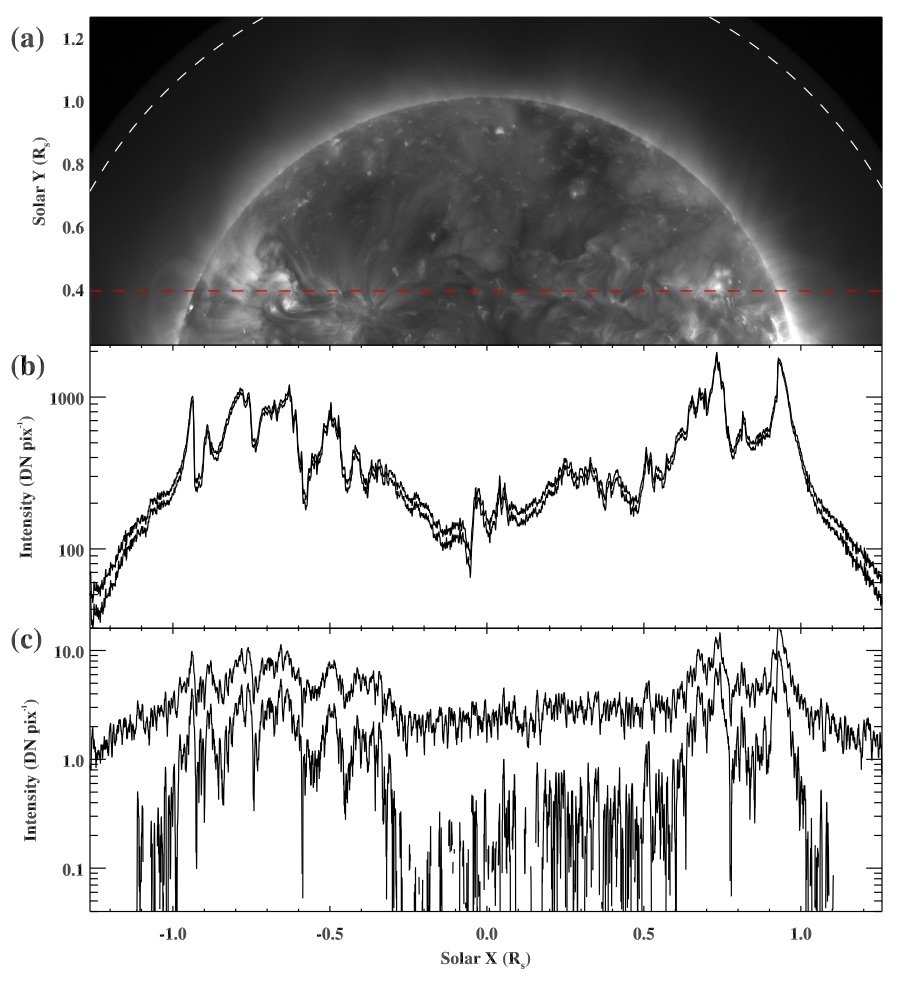}}
   \caption{(a) An AIA 193 channel image from 2015/01/01 03:00. The dashed white line shows a heliocentric height of 1.45\Rs, and the dashed red line shows a horizontal cut across the image. (b) The intensity along the dashed red line for the channel with the highest mean intensity (193), with the two lines showing the width of the measurmeent uncertainties. (c) As (b) for the channel with the lowest mean intensity (94).}
    \label{aiaerrors}
  \end{figure}

The uncertainty of the measurements given by the AIA Solarsoft routine aia\_bp\_estimate\_error includes the Poisson photon count, dark subtraction, read noise, count quantization and image compression uncertainties. Figures \ref{aiaerrors}b and c shows the range of intensities enclosed by the estimated errors for the 193 and 94 channel respectively, for a horizontal cut across the images shown by the dashed red line in figure \ref{aiaerrors}a. In high-signal regions/channels, the measurement error is small and the dominant uncertainty is in the response functions (calibration uncertainty). In low-signal regions/channels the method is influenced by both the response function and measurement uncertainties. At the expense of spatial and temporal resolution, rebinning images to smaller size through neighbourhood averaging, and combining two or more consecutive observations over time, will decrease measurement noise in the low signal channels to a more acceptable level.

\begin{figure}    
   \centerline{\includegraphics[width=0.98\textwidth,clip=]{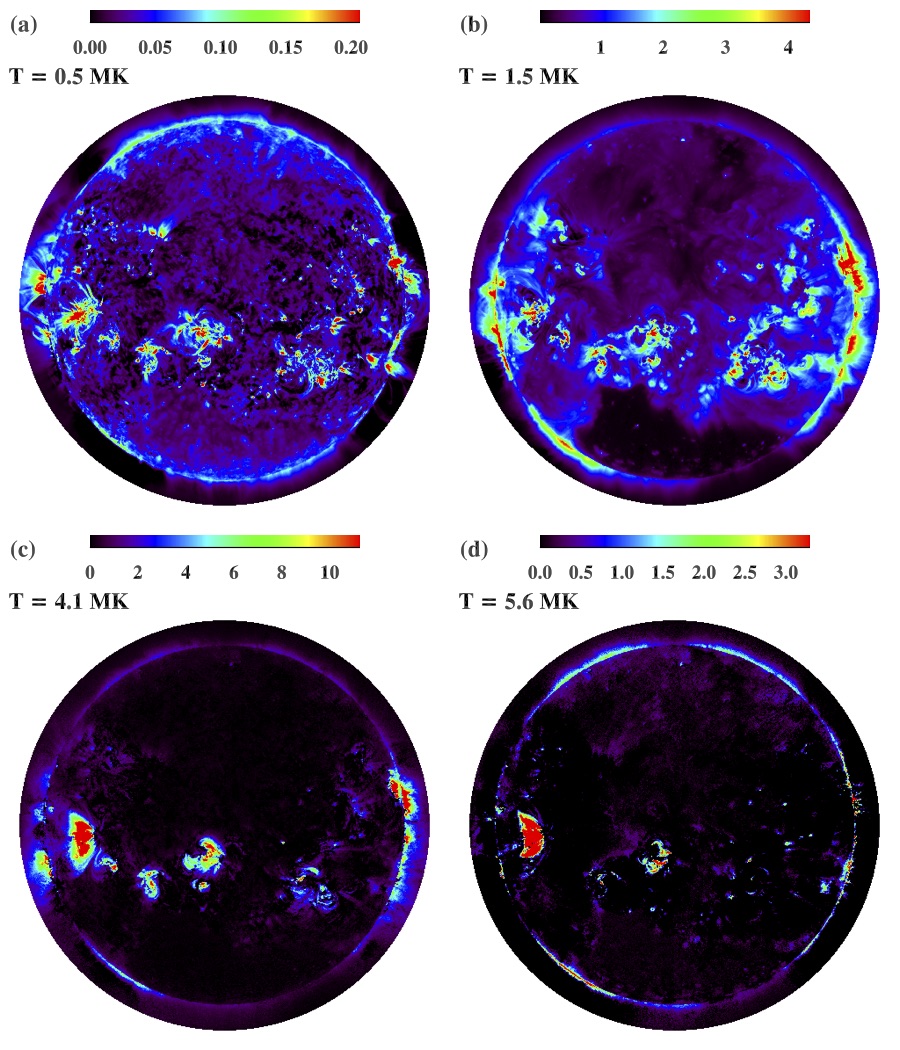}}
   \caption{Emission for four different temperatures as indicated in each panel. The field of view is curtailed to a maximum heliocentric distance of 1.15\Rs. The color bars give EM in units of $10^{26}$\EM.}
    \label{aiadem0}
  \end{figure}

The dataset of 2015/01/01 is rebinned to $512 \times 512$ pixels, from the original $4096 \times 4096$ pixels. Since 64 original measurements are combined (averaged) for each pixel, the measurement noise decreases by a factor of $1/8$. The DEM method is applied to all pixels at heights below 1.15\Rs, and DEMs converted to EM by product with the width of the temperature bins. Emission is shown for four example temperatures in figure \ref{aiadem0}. 

\begin{figure}    
   \centerline{\includegraphics[width=0.98\textwidth,clip=]{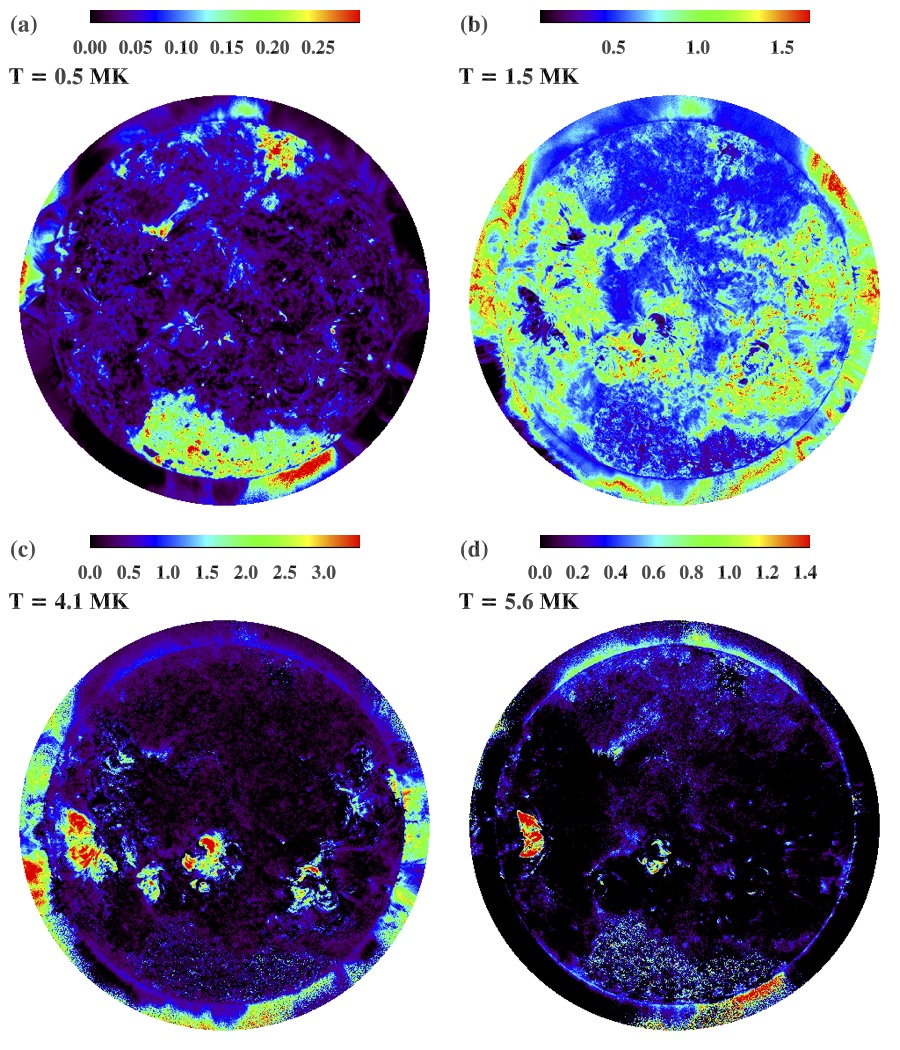}}
   \caption{Fractional emission (FEM) for four different temperatures as indicated in each panel. The field of view is curtailed to a maximum heliocentric distance of 1.15\Rs. The color bars give FEM in \%.}
    \label{aiadem1}
  \end{figure}

Effective visualisation of DEMs is challenging, since the output result from an imaging instrument is a datacube, thus one can show emission at a given temperature yet the context of emission at other temperatures is absent. Such direct DEM images are also dominated by the high emission, at all temperatures, of active regions. One effective method is the emission-weighted-mean or median temperature displayed with a colour/hue table that can show temperatures and emission, as shown for example in figure 15 of \opencite{plowman2013}. For visually comparing DEM maps in the context of dominance of different regions by certain temperature ranges, we introduce the simple concept of Fractional emission measure (FEM). FEM in a temperature bin (indexed $j$) is calculated from a DEM by
  \begin{equation}
  \label{fracDEM}
  FEM_j=\frac{DEM_j \Delta T}{\sum_j DEM_j \Delta T}, 
  \end{equation}
so the FEM in a given temperature bin gives the fraction of emission at that temperature compared to the total emission integrated over all temperatures. FEM maps are shown in figure \ref{aiadem1}. These maps, for regions on the disk, are a powerful visualisation of the different general temperature dependencies of large-scale coronal features:
\begin{itemize}
\item At T$=0.5$MK, the FEM maps are dominated strongly by coronal holes and filament channels. This is an effective way of identifying these regions. 
\item At T$=1.5$MK, broad regions of the quiet corona and coronal holes have high FEM. Quiet regions surrounding active regions are particularly strong. Note that active regions have generally very low FEM at this temperature.
\item At T$=4.1$MK, all regions except active regions have low FEM. Note in the original EM maps, that active regions have high EM at all temperatures compared to other regions due to their high mass. The FEM maps, through normalization by the total EM, removes this effect and shows that, despite the multithermality of active regions, their emission is dominated by high temperatures.
\item At T$=5.6$MK, only the hot cores of the large active regions have high FEM. The quiet coronal regions have close to zero FEM at this temperature.
\end{itemize}

The DEMs in off-limb regions are hard to interpret and are subject to the bias towards high temperatures with increasing height, given the large height scale for hot structures, as explained by, e.g. \opencite{aschwanden2005}. Solar rotational tomography offers a solution to this line-of-sight problem. A framework for tomography combined with a DEM analysis is given by \opencite{nuevo2015}, where the intensity from each channel, observed from several different viewpoints, is reconstructed in a 3D volume of emission, and a local DEM computed at each voxel. 

\section{Summary}
\label{summary}
A new DEM method is presented which is reasonably fast, simple in concept, and simple to implement. It performs well on tests involving model DEMs and synthetic data based on the AIA/SDO instrument. In particular, the correlation between the model input DEMs and SITES inversions is excellent for a broad range of coronal temperatures. SITES performs less well on very narrow DEM peaks, and performs very poorly for temperatures below $\sim$0.5MK. This weakness is likely due to the limitations of the AIA/SDO instrumental temperature response curves rather than the SITES inversion itself, since other inversion methods show the same failing.

Applied to a set of AIA/SDO observations of the full-disk corona, SITES gives sensible values of emission as a function of temperature. Fractional emission measure is introduced as a simple yet powerful method to visualise DEM results within images, enabling straightforward comparison of different temperature regimes between regions. 

The computational speed of the method compares well with most methods, but cannot compete with the sparse matrix approach of \opencite{cheung2015}. However, the main advantages of SITES is its simplicity of concept and application, and its non-subjectiveness. Equations \ref{dem1} and \ref{iexp} form the core of the iterative procedure, and are simple to implement. The results of any DEM inversion method are subject to choices of fitting parameters. In the case of SITES, there is only one parameters which effects the result - the width of the smoothing kernel. Thus the method is relatively non-subjective.

The incentive for developing the method is to analyse large datasets, thus enabling large-scale studies of coronal changes over long time-scales using AIA/SDO. The method has therefore not been tested on flare-like temperatures. Reliable studies of such high temperatures need measurements by other instruments, possibly in combination with AIA/SDO. Given a set of temperature response functions and error estimates, the method presented here should work reliably - this will be investigated in the near future.

Future work by the authors (paper in preparation) involves a gridding method that may be used with any DEM inversion method to increase computational efficiency by one or two orders of magnitude. This will enable rapid processing of large datasets for AIA/SDO and other current or future instruments. The software for the DEM fitting method of this paper, plus the FEM visualisation method, written in IDL, is available by email request to the authors. 

\begin{acks}
James Pickering is supported by an STFC studentship. Part of Huw Morgan's work on this project is supported by an STFC consolidated grant to Aberystwyth University. CHIANTI is a collaborative project involving George Mason University, the University of Michigan (USA), University of Cambridge (UK) and NASA Goddard Space Flight Center (USA). The AIA/SDO data is courtesy of NASA/SDO and the AIA science team.
\end{acks}

\bibliographystyle{spr-mp-sola}
\bibliography{./biblio.bib}

\begin{thebibliography}{17}
\ifx \bisbn   \undefined \def \bisbn  #1{ISBN #1}\fi
\ifx \binits  \undefined \def \binits#1{#1}\fi
\ifx \bauthor  \undefined \def \bauthor#1{#1}\fi
\ifx \batitle  \undefined \def \batitle#1{#1}\fi
\ifx \bjtitle  \undefined \def \bjtitle#1{\textit{#1}}\fi
\ifx \bvolume  \undefined \def \bvolume#1{\textbf{#1}}\fi
\ifx \byear  \undefined \def \byear#1{#1}\fi
\ifx \bissue  \undefined \def \bissue#1{#1}\fi
\ifx \bfpage  \undefined \def \bfpage#1{#1}\fi
\ifx \blpage  \undefined \def \blpage #1{#1}\fi
\ifx \burl  \undefined \def \burl#1{\textsf{#1}}\fi
\ifx \href  \undefined \def \href#1#2{\textsf{#2}}\fi
\ifx \doiurl  \undefined \def
  \doiurl#1{\href{http://dx.doi.org/#1}{\textsf{#1}}}\fi
\ifx \betal  \undefined \def \betal{\textit{et al.}}\fi
\ifx \binstitute  \undefined \def \binstitute#1{#1}\fi
\ifx \bctitle  \undefined \def \bctitle#1{#1}\fi
\ifx \beditor  \undefined \def \beditor#1{#1}\fi
\ifx \bpublisher  \undefined \def \bpublisher#1{#1}\fi
\ifx \bbtitle  \undefined \def \bbtitle#1{\textit{#1}}\fi
\ifx \bedition  \undefined \def \bedition#1{#1}\fi
\ifx \bseriesno  \undefined \def \bseriesno#1{\textbf{#1}}\fi
\ifx \blocation  \undefined \def \blocation#1{#1}\fi
\ifx \bsertitle  \undefined \def \bsertitle#1{\textit{#1}}\fi
\ifx \bsnm \undefined \def \bsnm#1{#1}\fi
\ifx \bsuffix \undefined \def \bsuffix#1{#1}\fi
\ifx \bparticle \undefined \def \bparticle#1{#1}\fi
\ifx \barticle \undefined \def \barticle#1{}\fi
\ifx \botherref \undefined \def \botherref#1{}\fi
\ifx \url \undefined \def \url#1{\textsf{#1}}\fi
\ifx \bchapter \undefined \def \bchapter#1{}\fi
\ifx \bbook \undefined \def \bbook#1{}\fi
\ifx \bcomment \undefined \def \bcomment#1{#1}\fi
\ifx \oauthor \undefined \def \oauthor#1{#1}\fi
\ifx \citeauthoryear \undefined \def \citeauthoryear#1{#1}\fi
\def \endbibitem {}
\ifx \bconflocation  \undefined \def \bconflocation#1{#1} \fi

\bibitem[\protect\citeauthoryear{{Aschwanden}}{2005}]{aschwanden2005}
\begin{bbook}
\bauthor{\bsnm{{Aschwanden}}, \binits{M.J.}}:
\byear{2005},
\bbtitle{{Physics of the Solar Corona. An Introduction with Problems and
  Solutions (2nd edition)}}.
\end{bbook}
\endbibitem

\bibitem[\protect\citeauthoryear{Boerner \textit{et~al.}}{2014}]{boerner2014}
\begin{barticle}
\bauthor{\bsnm{Boerner}, \binits{P.F.}},
\bauthor{\bsnm{Testa}, \binits{P.}},
\bauthor{\bsnm{Warren}, \binits{H.}},
\bauthor{\bsnm{Weber}, \binits{M.A.}},
\bauthor{\bsnm{Schrijver}, \binits{C.J.}}:
\byear{2014},
\batitle{Photometric and thermal cross-calibration of solar euv instruments}.
\bjtitle{Solar Physics}
\bvolume{289}(\bissue{6}),
\bfpage{2377}\,--\,\blpage{2397}.
doi:\doiurl{10.1007/s11207-013-0452-z}.
\burl{https://doi.org/10.1007/s11207-013-0452-z}.
\end{barticle}
\endbibitem

\bibitem[\protect\citeauthoryear{{Cheung} \textit{et~al.}}{2015}]{cheung2015}
\begin{barticle}
\bauthor{\bsnm{{Cheung}}, \binits{M.C.M.}},
\bauthor{\bsnm{{Boerner}}, \binits{P.}},
\bauthor{\bsnm{{Schrijver}}, \binits{C.J.}},
\bauthor{\bsnm{{Testa}}, \binits{P.}},
\bauthor{\bsnm{{Chen}}, \binits{F.}},
\bauthor{\bsnm{{Peter}}, \binits{H.}},
\bauthor{\bsnm{{Malanushenko}}, \binits{A.}}:
\byear{2015},
\batitle{{Thermal Diagnostics with the Atmospheric Imaging Assembly on board
  the Solar Dynamics Observatory: A Validated Method for Differential Emission
  Measure Inversions}}.
\bjtitle{\apj}
\bvolume{807},
\bfpage{143}.
doi:\doiurl{10.1088/0004-637X/807/2/143}.
\end{barticle}
\endbibitem

\bibitem[\protect\citeauthoryear{{Del Zanna}}{2013}]{delzanna2013}
\begin{barticle}
\bauthor{\bsnm{{Del Zanna}}, \binits{G.}}:
\byear{2013},
\batitle{{The multi-thermal emission in solar active regions}}.
\bjtitle{\aap}
\bvolume{558},
\bfpage{A73}.
doi:\doiurl{10.1051/0004-6361/201321653}.
\end{barticle}
\endbibitem

\bibitem[\protect\citeauthoryear{{Dere} \textit{et~al.}}{1997}]{dere1997}
\begin{barticle}
\bauthor{\bsnm{{Dere}}, \binits{K.P.}},
\bauthor{\bsnm{{Landi}}, \binits{E.}},
\bauthor{\bsnm{{Mason}}, \binits{H.E.}},
\bauthor{\bsnm{{Monsignori Fossi}}, \binits{B.C.}},
\bauthor{\bsnm{{Young}}, \binits{P.R.}}:
\byear{1997},
\batitle{{CHIANTI - an atomic database for emission lines}}.
\bjtitle{\aaps}
\bvolume{125},
\bfpage{149}\,--\,\blpage{173}.
doi:\doiurl{10.1051/aas:1997368}.
\end{barticle}
\endbibitem

\bibitem[\protect\citeauthoryear{{Druckm{\"u}ller}}{2009}]{druckmuller2009}
\begin{barticle}
\bauthor{\bsnm{{Druckm{\"u}ller}}, \binits{M.}}:
\byear{2009},
\batitle{{Phase Correlation Method for the Alignment of Total Solar Eclipse
  Images}}.
\bjtitle{\apj}
\bvolume{706},
\bfpage{1605}\,--\,\blpage{1608}.
doi:\doiurl{10.1088/0004-637X/706/2/1605}.
\end{barticle}
\endbibitem

\bibitem[\protect\citeauthoryear{{Fisher} and {Welsch}}{2008}]{fisher2008}
\begin{bchapter}
\bauthor{\bsnm{{Fisher}}, \binits{G.H.}},
\bauthor{\bsnm{{Welsch}}, \binits{B.T.}}:
\byear{2008},
\bctitle{{FLCT: A Fast, Efficient Method for Performing Local Correlation
  Tracking}}.
In: \beditor{\bsnm{{Howe}}, \binits{R.}},
\beditor{\bsnm{{Komm}}, \binits{R.W.}},
\beditor{\bsnm{{Balasubramaniam}}, \binits{K.S.}},
\beditor{\bsnm{{Petrie}}, \binits{G.J.D.}} (eds.)
\bbtitle{Subsurface and Atmospheric Influences on Solar Activity},
\bsertitle{Astronomical Society of the Pacific Conference Series}
\bseriesno{383},
\bfpage{373}.
\end{bchapter}
\endbibitem

\bibitem[\protect\citeauthoryear{{Hahn} and {Savin}}{2014}]{hahn2014}
\begin{barticle}
\bauthor{\bsnm{{Hahn}}, \binits{M.}},
\bauthor{\bsnm{{Savin}}, \binits{D.W.}}:
\byear{2014},
\batitle{{Evidence for Wave Heating of the Quiet-Sun Corona}}.
\bjtitle{\apj}
\bvolume{795},
\bfpage{111}.
doi:\doiurl{10.1088/0004-637X/795/2/111}.
\end{barticle}
\endbibitem

\bibitem[\protect\citeauthoryear{{Hahn}, {Landi}, and {Savin}}{2011}]{hahn2011}
\begin{barticle}
\bauthor{\bsnm{{Hahn}}, \binits{M.}},
\bauthor{\bsnm{{Landi}}, \binits{E.}},
\bauthor{\bsnm{{Savin}}, \binits{D.W.}}:
\byear{2011},
\batitle{{Differential Emission Measure Analysis of a Polar Coronal Hole during
  the Solar Minimum in 2007}}.
\bjtitle{\apj}
\bvolume{736},
\bfpage{101}.
doi:\doiurl{10.1088/0004-637X/736/2/101}.
\end{barticle}
\endbibitem

\bibitem[\protect\citeauthoryear{{Hannah} and {Kontar}}{2012}]{hannah2012}
\begin{barticle}
\bauthor{\bsnm{{Hannah}}, \binits{I.G.}},
\bauthor{\bsnm{{Kontar}}, \binits{E.P.}}:
\byear{2012},
\batitle{{Differential emission measures from the regularized inversion of
  Hinode and SDO data}}.
\bjtitle{\aap}
\bvolume{539},
\bfpage{A146}.
doi:\doiurl{10.1051/0004-6361/201117576}.
\end{barticle}
\endbibitem

\bibitem[\protect\citeauthoryear{{Landi} \textit{et~al.}}{2012}]{landi2012}
\begin{barticle}
\bauthor{\bsnm{{Landi}}, \binits{E.}},
\bauthor{\bsnm{{Del Zanna}}, \binits{G.}},
\bauthor{\bsnm{{Young}}, \binits{P.R.}},
\bauthor{\bsnm{{Dere}}, \binits{K.P.}},
\bauthor{\bsnm{{Mason}}, \binits{H.E.}}:
\byear{2012},
\batitle{{CHIANTI - An Atomic Database for Emission Lines. XII. Version 7 of
  the Database}}.
\bjtitle{\apj}
\bvolume{744},
\bfpage{99}.
doi:\doiurl{10.1088/0004-637X/744/2/99}.
\end{barticle}
\endbibitem

\bibitem[\protect\citeauthoryear{{Mackovjak}, {Dzif{\v c}{\'a}kov{\'a}}, and
  {Dud{\'{\i}}k}}{2014}]{mackovjak2014}
\begin{barticle}
\bauthor{\bsnm{{Mackovjak}}, \binits{{\v S}.}},
\bauthor{\bsnm{{Dzif{\v c}{\'a}kov{\'a}}}, \binits{E.}},
\bauthor{\bsnm{{Dud{\'{\i}}k}}, \binits{J.}}:
\byear{2014},
\batitle{{Differential emission measure analysis of active region cores and
  quiet Sun for the non-Maxwellian {$\kappa$}-distributions}}.
\bjtitle{\aap}
\bvolume{564},
\bfpage{A130}.
doi:\doiurl{10.1051/0004-6361/201323054}.
\end{barticle}
\endbibitem

\bibitem[\protect\citeauthoryear{{Morgan} and {Taroyan}}{2017}]{morgan2017}
\begin{barticle}
\bauthor{\bsnm{{Morgan}}, \binits{H.}},
\bauthor{\bsnm{{Taroyan}}, \binits{Y.}}:
\byear{2017},
\batitle{Global conditions in the solar corona from 2010 to 2017}.
\bjtitle{Science Advances}
\bvolume{3}(\bissue{7}).
doi:\doiurl{10.1126/sciadv.1602056}.
\burl{https://advances.sciencemag.org/content/3/7/e1602056}.
\end{barticle}
\endbibitem

\bibitem[\protect\citeauthoryear{Morgan and Druckm{\"u}ller}{2014}]{morgan2014}
\begin{barticle}
\bauthor{\bsnm{Morgan}, \binits{H.}},
\bauthor{\bsnm{Druckm{\"u}ller}, \binits{M.}}:
\byear{2014},
\batitle{Multi-scale gaussian normalization for solar image processing}.
\bjtitle{Solar Physics}
\bvolume{289}(\bissue{8}),
\bfpage{2945}\,--\,\blpage{2955}.
doi:\doiurl{10.1007/s11207-014-0523-9}.
\burl{https://doi.org/10.1007/s11207-014-0523-9}.
\end{barticle}
\endbibitem

\bibitem[\protect\citeauthoryear{Nuevo \textit{et~al.}}{2015}]{nuevo2015}
\begin{barticle}
\bauthor{\bsnm{Nuevo}, \binits{F.A.}},
\bauthor{\bsnm{V{\'{a}}squez}, \binits{A.M.}},
\bauthor{\bsnm{Landi}, \binits{E.}},
\bauthor{\bsnm{Frazin}, \binits{R.}}:
\byear{2015},
\batitle{{MULTIMODAL} {DIFFERENTIAL} {EMISSION} {MEASURE} {IN} {THE} {SOLAR}
  {CORONA}}.
\bjtitle{The Astrophysical Journal}
\bvolume{811}(\bissue{2}),
\bfpage{128}.
doi:\doiurl{10.1088/0004-637X/811/2/128}.
\burl{https://doi.org/10.1088/0004-637x/811/2/128}.
\end{barticle}
\endbibitem

\bibitem[\protect\citeauthoryear{{Plowman}, {Kankelborg}, and
  {Martens}}{2013}]{plowman2013}
\begin{barticle}
\bauthor{\bsnm{{Plowman}}, \binits{J.}},
\bauthor{\bsnm{{Kankelborg}}, \binits{C.}},
\bauthor{\bsnm{{Martens}}, \binits{P.}}:
\byear{2013},
\batitle{{Fast Differential Emission Measure Inversion of Solar Coronal Data}}.
\bjtitle{\apj}
\bvolume{771},
\bfpage{2}.
doi:\doiurl{10.1088/0004-637X/771/1/2}.
\end{barticle}
\endbibitem

\bibitem[\protect\citeauthoryear{{Weber} \textit{et~al.}}{2005}]{weber2005}
\begin{barticle}
\bauthor{\bsnm{{Weber}}, \binits{M.A.}},
\bauthor{\bsnm{{Schmelz}}, \binits{J.T.}},
\bauthor{\bsnm{{DeLuca}}, \binits{E.E.}},
\bauthor{\bsnm{{Roames}}, \binits{J.K.}}:
\byear{2005},
\batitle{{Isothermal Bias of the ``Filter Ratio'' Method for Observations of
  Multithermal Plasma}}.
\bjtitle{\apjl}
\bvolume{635},
\bfpage{101}\,--\,\blpage{104}.
doi:\doiurl{10.1086/499125}.
\end{barticle}
\endbibitem

\end{thebibliography}

\end{article} 

\end{document}